\documentclass[apj]{emulateapj}

\newcommand{\bgq}{B$_{gq}$}
\newcommand{\bgg}{B$_{gg}$}
\newcommand{\agq}{A$_{gq}$}
\newcommand{\kms}{km s$^{-1}$}
\newcommand{\hinv}{$h^{-1}$}

\begin{document}

\title{Wide Field Multiband Imaging of Low Redshift Quasar Environments}

\author{
Jennifer E. Scott\altaffilmark{1},
Alireza Rafiee\altaffilmark{1,2},
Jill Bechtold\altaffilmark{3},
Erica Ellingson\altaffilmark{4},
Christopher Thibodeau\altaffilmark{1,5},
Michael Richmond\altaffilmark{1,6}}

\altaffiltext{1}{
Department of Physics, Astronomy and Geosciences,
Towson University, 8000 York Road,
Towson, MD 21252-0001;
[jescott,arafiee]@towson.edu}

\altaffiltext{2}{current address:
Department of Physics and Astronomy,
York University,
4700 Keele Street, Toronto, Ontario, M3J 1P3, Canada;arafiee@yorku.ca
}

\altaffiltext{3}{
Steward Observatory,
University of Arizona, 933 North Cherry Ave., Tucson, AZ  85721-0065;
jbechtold@as.arizona.edu
}

\altaffiltext{4}{
Center for Astrophysics and Space Astronomy,
389 UCB,
University of Colorado, Boulder, CO  80309-0389;
erica.ellingson@casa.colorado.edu
}

\altaffiltext{5}{current address:
Department of Astronomy,
University of Massachusetts,
710 North Pleasant Street, Amherst, MA 01003-9305;crthibodeau@astro.umass.edu 
}

\altaffiltext{6}{current address:
Space and Naval Warfare Systems Command Atlantic,
P.O. Box 190022,
North Charleston, SC 29419-9022
}

\addtocounter{footnote}{-10}

\begin{abstract}
We present photometry of the large scale environments of a sample of twelve
broad line AGN with $0.06 < z < 0.37$
from deep images in the SDSS
$u$, $g$, $r$, and $i$ filters taken with the 90Prime prime focus camera on the 
Steward Observatory Bok Telescope.
We measure galaxy clustering around these AGN using two standard techniques: 
correlation amplitude (\bgq) and the two point correlation function. 
We find average correlation amplitudes for the 10 radio quiet objects in the sample
equal to (9$\pm$18, 144$\pm$114, -39$\pm$56, 295$\pm$260) Mpc$^{1.77}$ in 
($u$, $g$, $r$, $i$), all consistent with the expectation from galaxy clustering. 
Using a ratio of the galaxy-quasar cross-correlation function to the galaxy 
autocorrelation function, we calculate the relative bias of galaxies and AGN, $b_{gq}$. 
The bias in the $u$ band, $b_{gq}=3.08\pm0.51$ is larger compared to that calculated in the other bands, but
it does not 
correlate with AGN luminosity, black hole mass, or AGN 
activity via the luminosity of the [\ion{O}{3}] emission line.  Thus ongoing nuclear accretion 
activity is not reflected in the large scale environments from $\sim$10 \hinv\ kpc to $\sim$0.5 \hinv\ Mpc and may 
indicate a 
non-merger mode of AGN activity and/or a significant delay between galaxy mergers and nuclear activity 
in this sample of mostly radio quiet quasars.
\end{abstract}


\section{Introduction}
In local galaxies, the masses of the nuclear black holes are correlated
with the stellar velocity dispersion of the host galaxy
bulges, the M$_{\rm BH}$-$\sigma$ relation \citep{FerrareseMerritt2000,Gebhardt2000,Tremaine2002,McLureDunlop2002,
MarconiHunt2003}.
The host galaxies of low redshift ($z$$<$0.4) quasars 
are consistent with these correlations, with the most powerful quasars typically
inhabiting massive, early-type galaxies \citep{McLeod1995,McLeod1999,McLure1999,Hamilton2002,Dunlop2003,Floyd2004}.

Theoretical modeling establishes a framework for the coevolution of galaxies and supermassive black holes
such that after the initial formation of
central black holes in galaxies, nuclear activity and star formation
are triggered by a merger event
\citep{Barnes1992,Kauffmann2000,Volonteri2003,Wyithe2003,Granato2004,Haiman2004,
DiMatteo2005,Springel2005a,Springel2005b,Croton2006,Hopkins2006a,Hopkins2006b,Hopkins2008,Somerville2008}.
These hierarchical models
make specific predictions for the large scale environments of luminous
quasars.

Observational studies have investigated the wide parameter space of this problem 
by examining: the quasar-quasar and galaxy-quasar clustering in terms of correlation length and amplitude;
nearest neighbor and other statistics for estimating environment density;
properties of galaxies in AGN environments; and the properties of AGN host galaxies,
covering also a range in redshift and intrinsic AGN properties.
Many of these studies have used Sloan Digital Sky Survey (SDSS) data, thus concentrating on the $z \lesssim 0.3$ universe.

From the observational results, two points of consensus have emerged that generally agree with the models:
(1) AGN activity is associated with galaxy mergers and enhanced star formation in their hosts \citep{Sanders1988,Kauffmann2004,
Guyon2006,Canalizo2007,Bennert2008,Coil2009,
Veilleux2009,Teng2010},
although there is increasing evidence, e.g. due to lack of correlation of galaxy structural parameters with local environment density,
\citep{Kauffmann2004} and an absence of increased rate of disturbed host morphologies in AGN hosts up to $z\sim1.3$,\citep{Grogin2005,Gabor2009},
that the onset of nuclear activity is delayed with respect to the initial
galaxy interaction \citep{Li2006,Li2008,Ellison2013} and that there is a significant non-merger mode to AGN activity \citep{HopkinsHernquist2006};
(2) on megaparsec scales, bright AGN cluster like L$^{*}$ galaxies \citep{Adelberger2005,Li2006,Serber2006,Coldwell2006,Coil2007,Gilli2009}, 
and over time, they occupy dark matter halos of 
constant mass ($\sim2-3 \times 10^{12}$ \hinv\ M$_{\odot}$) so they were more biased tracers of mass at higher redshift
than they are in the current epoch
\citep{Porciani2004,Wake2004,Croom2005,Myers2006,Myers2007a,Shen2007,Ross2009}.

Quasar-quasar clustering shows little or no dependence on quasar luminosity 
\citep{Croom2005,Porciani2006,Myers2006,Myers2007a,Shen2009} 
except at the lowest redshifts \citep{Wake2004,Constantin2006}.
Galaxy clustering around AGN does correlate with nuclear activity
especially for luminous, strongly accreting AGN, which tend to lie in 
massive hosts ($M^{*} \gtrsim 3 \times 10^{10} M_{\odot}$) with higher than average SFR \citep{Kauffmann2003},
and are more likely to
be found in lower density environments \citep{Kauffmann2004,Constantin2008,Silverman2009,Sabater2012,Sabater2013}.
No quasar luminosity dependence on galaxy-quasar clustering is seen in the DEEP2 sample with $z\sim1$ \citep{Coil2007}, though
quasars are found to cluster like blue, star-forming galaxies.

AGN fueling via a merger-independent quiescent or Seyfert mode 
\citep{HopkinsHernquist2006} may be most important for weaker AGN, as they
show no preference for low density regions in low redshift samples \citep{Kauffmann2004}
and X-ray selected AGN at $z\sim1$ with lower mass hosts ($M^{*} \lesssim  10^{11} M_{\odot}$)
which are found in a broad range of environments \citep{Silverman2009}.
However, there is evidence that 
the transition from Seyferts to LINERS proposed by \cite{Kewley2006} is 
affected by the density of the environment, 
in SDSS samples at low redshift \citep{Constantin2008} and in the X-ray selected $z\sim1$ AGN sample of \cite{MonteroDorta2009}.
These authors, and \cite{Constantin2006} also find that 
that the  LINERS cluster more strongly than Seyfert galaxies.

Unlike galaxies, quasars do show excess clustering  on small scales (10-100 \hinv\ kpc) up to $z\sim3$ \citep{Hennawi2006,Myers2007b,Shen2010}.
And while studies of the $\sim$1 \hinv\ Mpc environments around low redshift AGN show no distinct correlation between nuclear activity and the presence of 
galaxy neighbors on these scales \citep{Serber2006,Coldwell2006,Li2008}, there is evidence that the presence of a galaxy companion on smaller scales 
does influence AGN activity \citep{Alonso2007,Sabater2013}.
On larger scales,  
$z\gtrsim1$ radio loud quasars are found in richer environments than radio quiet objects \citep{Hall1998,Teplitz1999,Wold2003}
while for low redshift this distinction is less pronounced and 
AGN generally are not preferentially found at the 
centers of rich clusters \citep{Fisher1996,CroomShanks1999,McLure2001,Wold2001,Coldwell2002,Barr2003,Coldwell2006}. For cold, or QSO, mode 
accretion, this can be understood as inhibition due to
the stripping and harassment that occurs in these environments.
However, some studies have shown that quasars do trace the large scale structure of
galaxy clusters  \citep{Sochting2002,Sochting2004,Gilli2003,Silverman2008} on scales up to 10 \hinv\ Mpc.

In this paper, we investigate the large scale environments of a sample of 12 relatively bright low redshift ($0.06 < z < 0.37$) 
broad line AGN
using deep images in the SDSS $u$, $g$, $r$, and $i$ filters.  
The fields presented here overlap almost entirely with the SDSS, but our photometry goes significantly deeper, in some fields
and filters by up to 2.5-3 magnitudes. Our wide field multiband study on scales $\sim$1 degree, is distinct from previous single filter 
studies with {\it Hubble Space Telescope (HST)} using the $\sim$3 arcminute field of the Wide Field Planetary Camera 2  
\citep{Fisher1996,Finn2001,McLure2001}.
Likewise, the multiband nature of our study extends previous
wider field studies using the B$_{\rm J}$ and R data from
UK Schmidt plates  \citep{Brown2001,Coldwell2002,Sochting2002}.
This study is a blend of a 
statistical approach and a detailed study of each field. 
We calculate the correlation lengths and amplitudes of the galaxy-quasar clustering and 
compare results in different bands.
We also use our data and results from the literature to discuss each quasar's environment in detail.
Our primary aim is to probe galaxies as far down the luminosity function as allowed by the depth of the photometry to as wide an area around each quasar as possible
in order to test if galaxy clustering around quasars is comparable to that of galaxy-galaxy clustering to these depths at these scales.
Throughout this paper we assume a cosmology with $h=0.71$ ($H_{0}=100$ h \kms\ Mpc$^{-1}$), $\Omega_{m}=0.27$, and $\Omega_{\Lambda}=0.73$.

\section{Data}
\subsection{Observations and Reductions}

We made our observations with the 90Prime prime focus imager \citep{Williams2004} 
on the 2.3-meter Bok Telescope at Kitt Peak National Observatory in January - June, 2008. 
90Prime is a wide field camera consisting of four CCDs arranged in a square mosaic located at the prime focus of the telescope with a
field of view of $\sim$1 degree. We used the SDSS $u$, $g$, $r$, and $i$ filters for the photometry, and with the exception of HS0624+690, 
all our fields 
lie inside the SDSS footprint.  However, our photometry is significantly deeper than that of the SDSS survey, as discussed further in the next section.
We chose the quasars in our sample to have galaxy environments accessible to deep imaging with our 2-meter class telescope and with ultraviolet 
spectra in the archives of the  {\it Hubble Space Telescope} or {\it Far Ultraviolet Spectroscopic Explorer} for future studies of line of sight and
associated absorption in these systems. The lower bound on the
redshifts of the sample quasars was determined by the size of the chips in the CCD mosaic and the $\sim$Mpc scales targeted in this study.
We compiled a sample of $\sim$20 target fields for the first semester of the 2008 observing season and completed the photometry in at least two bands for
12 of those fields. Total exposure times were estimated for each field and filter in order to achieve significant detections of galaxies with M$^{*} + 2$ at
redshift of the quasar in each filter.
The final sample presented here includes one Seyfert, 9 radio quiet quasars and two radio loud quasars and so somewhat overrepresents the
radio loud population, although we separate these fields out in the discussion of our results.  
Follow up observations to obtain data on other fields of quasars with available UV spectra and with appropriate redshifts are possible.

The quasars themselves were centered on one of the chips of the mosaic, usually chip 1, rather than in the center of the field due the 
8.3\arcmin\ interchip gaps and 
several traps and other quality issues with the other chips in the mosaic at the time of the observations.
The raw data frames are multiextension FITS files consisting of eight 2048$\times$4096 images, one for each of the two amplifiers on each chip. 
Amplifiers 1 and 2 lie on chip 1, 3 and 4 on chip 2, 5 and 6 on chip 3, and amps 7 and 8 are on chip 4.
We give a summary of the observations in Table~\ref{tab-data}.

We reduced the data with standard techniques for bias, zero, and flat field corrections, including 
fringe corrections in the $i$ band, using the standard reduction tasks in the IRAF {\tt mosaic} package.  
\clearpage
\begin{turnpage}
\begin{deluxetable}{lccccccccc}
\tabletypesize{\scriptsize}
\tablecaption{90Prime Observations
\label{tab-data}}
\tablewidth{0pt}
\tablehead{
\colhead{QSO} &\colhead{Type\tablenotemark{1}} &\colhead{RA} &\colhead{Dec} &\colhead{Redshift} &\colhead{$r$}  &\colhead{$u$ exp.(s)}  &\colhead{$g$ exp.(s)} &\colhead{$r$ exp.(s)} &\colhead{$i$ exp.(s)}
}
\startdata
MRK586	    &Sy1.2   &02:07:49.8	 &+02:42:55	 &0.156	 &15.42                  &2800	&4400	&3600	&\nodata  \\
HS0624+690  &QSO     &06:30:02.5	 &+69:05:04	 &0.370	 &14.73\tablenotemark{2} &9200	&6400	&4400	&1200	  \\
PG0844+349  &Sy1.0   &08:47:42.4	 &+34:45:04      &0.064  &14.48                  &4400  &\nodata &7600  &\nodata  \\
PG0923+201  &Sy1.0   &09:25:54.7	 &+19:54:05	 &0.192	 &15.55	                 &5600	&5200	&2000	&2400	  \\
PG0953+414  &Sy1.0   &09:56:52.3	 &+41:15:22	 &0.234	 &14.93	                 &8000	&2400	&3600	&4200	   \\
PG1116+215  &Sy1.0   &11:19:08.7	 &+21:19:18	 &0.177	 &14.46	                 &5370	&400	&400	&2400	 \\
PG1307+085  &Sy1.2   &13:09:47.0	 &+08:19:48	 &0.155	 &15.59	                 &4400	&2000	&3200	&2400	 \\
PG1404+226  &NLS1    &14:06:21.9	 &+22:23:46	 &0.098	 &16.02	                 &5200	&4800	&6800	&2400	 \\
PG1444+407  &Sy1.0   &14:46:45.9	 &+40:35:06	 &0.267	 &15.81	                 &4400	&4000	&3200	&2400	 \\
PG1545+210  &Sy1.2   &15:47:43.5	 &+20:52:17	 &0.264	 &15.56	                 &6000	&4800	&2800	&2400	 \\
PG1612+261  &Sy1.5   &16:14:13.2	 &+26:04:16	 &0.131	 &15.87	                 &4000	&4800	&3800	&2400	 \\ 
Q2141+175   &Sy1.0   &21:43:35.5	 &+17:43:49	 &0.211	 &16.04	                 &\nodata &4500	&3000	&1500	 \\
\enddata
\tablenotetext{1}{\cite{VeronCetty2010}}
\tablenotetext{2}{Measured from 90Prime data}
\end{deluxetable}
\end{turnpage}
\clearpage

\subsection{Photometric Calibration and Catalog Construction}

\subsubsection{Calibration and Masking}
\begin{figure}
\epsscale{0.4}
\plotone{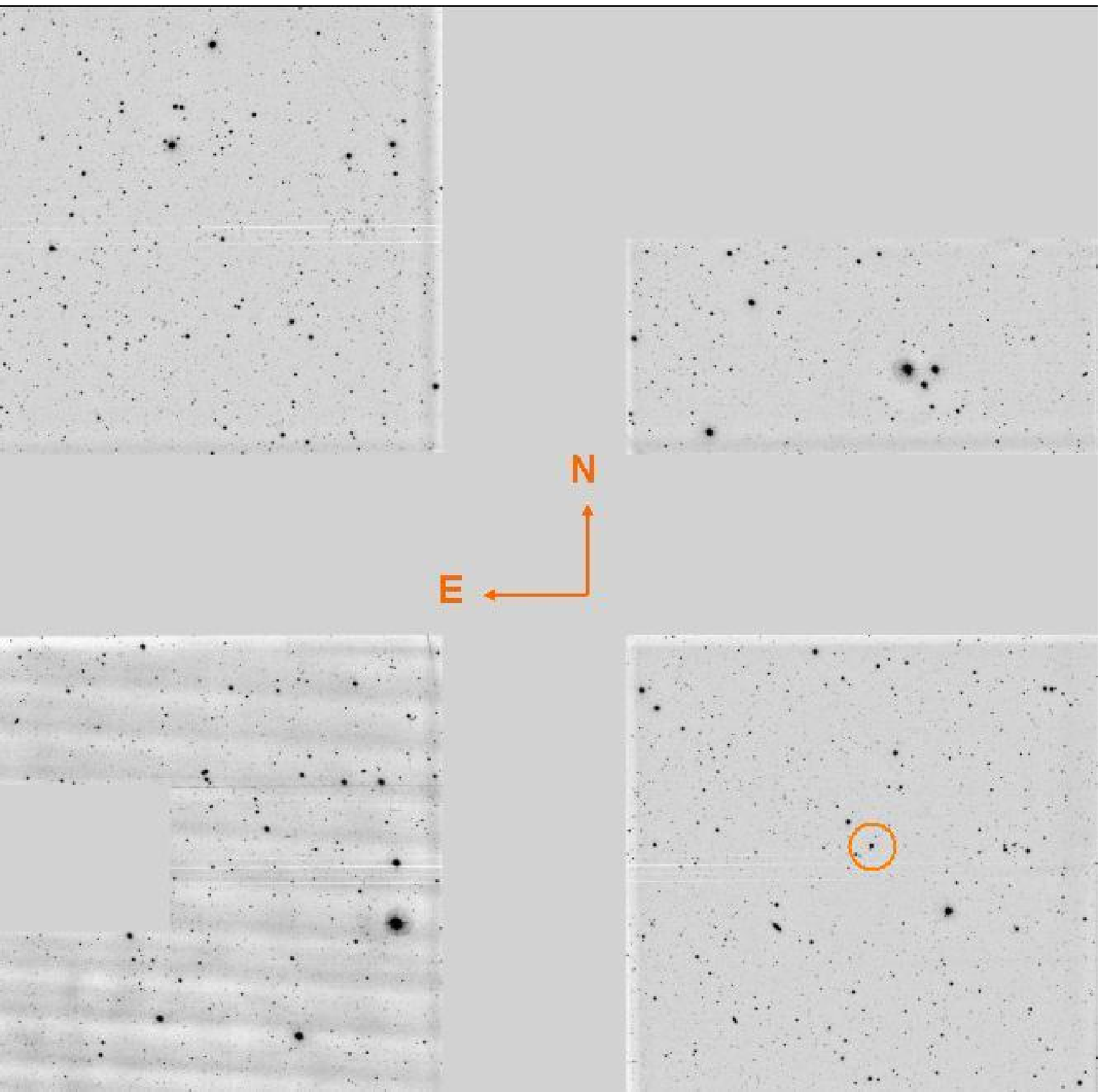}
\plotone{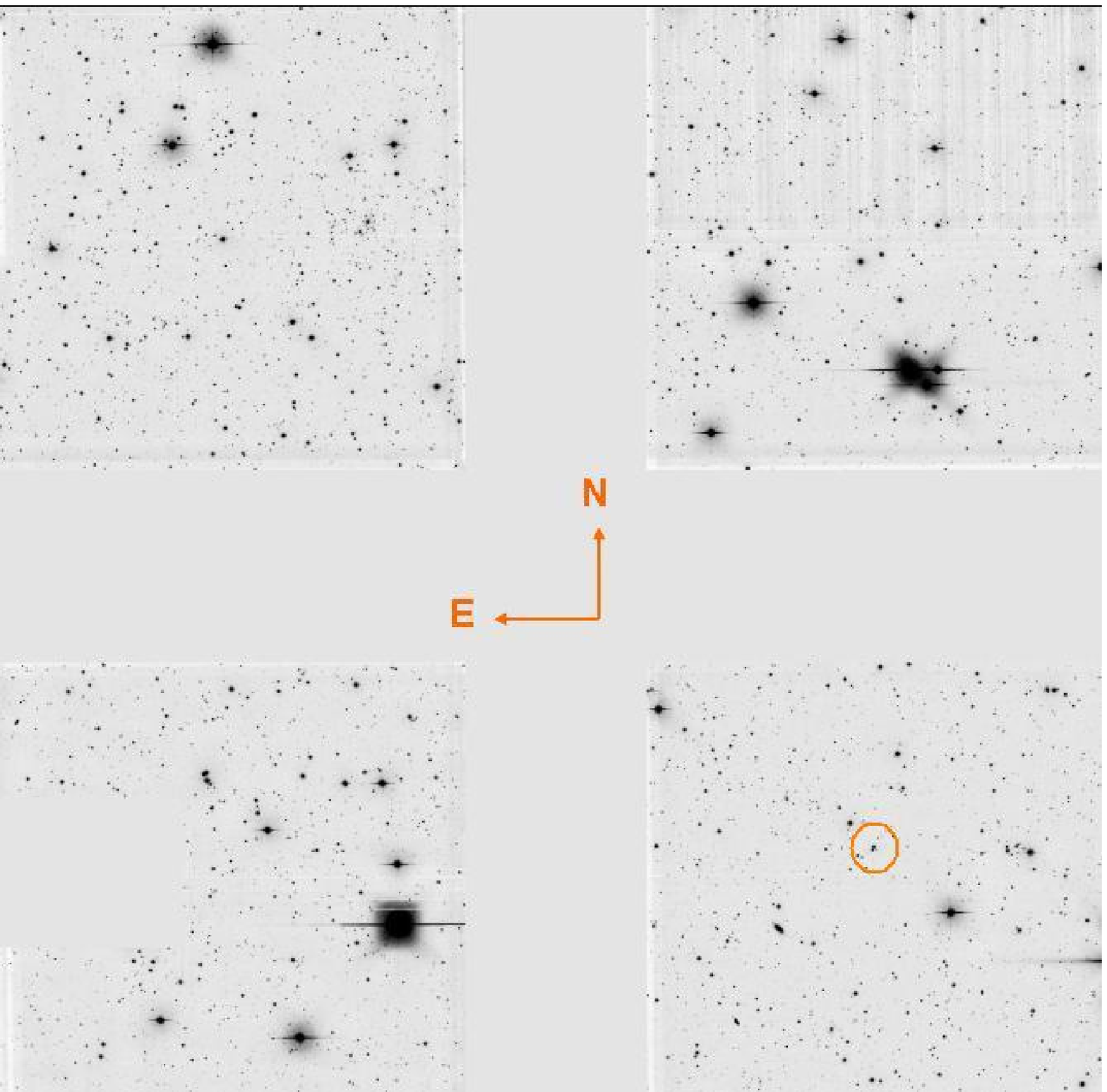}
\plotone{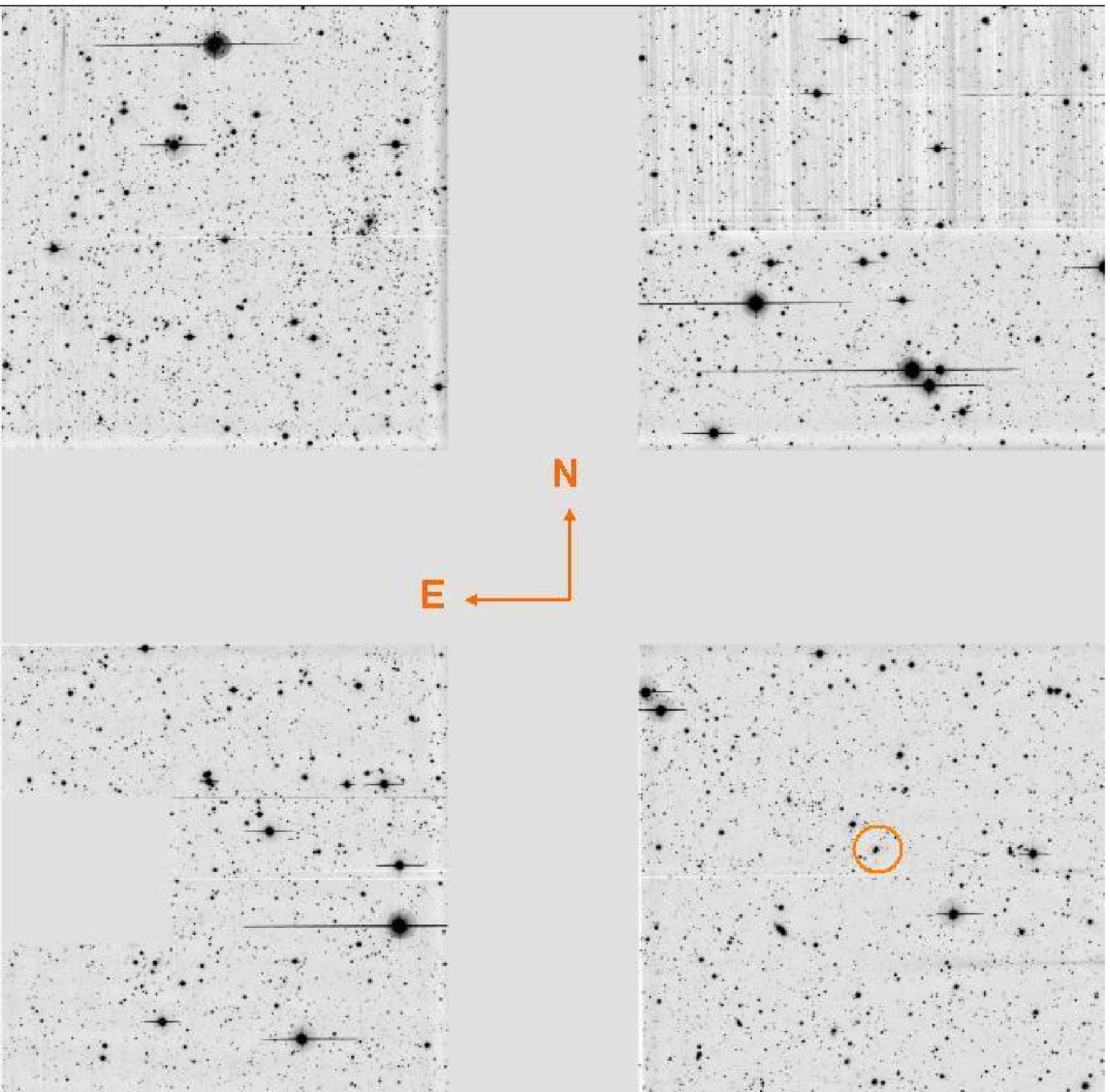}
\plotone{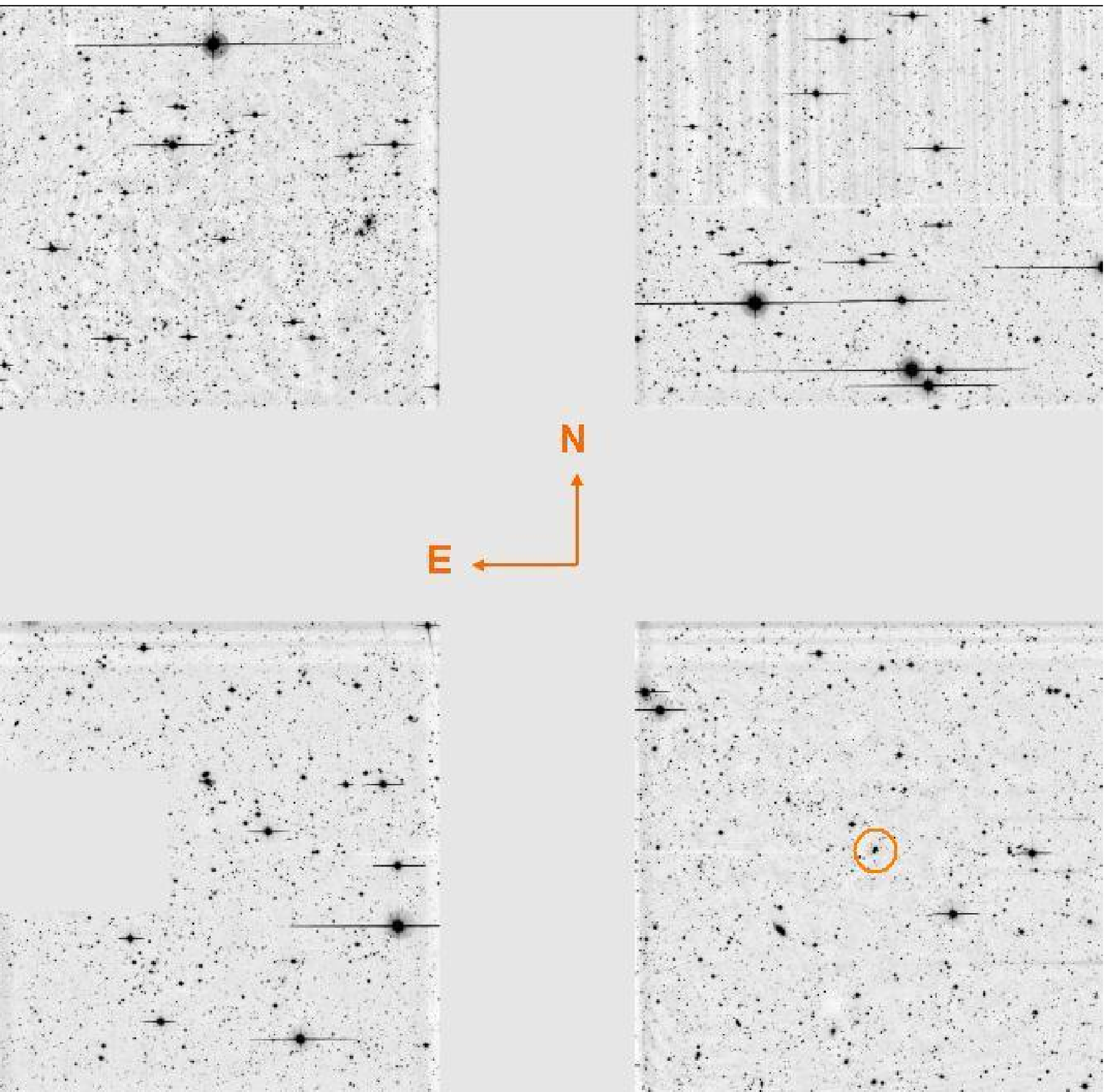}
\caption{
Example 90Prime data frame. The PG0923+201 field in 
$u$,$g$ (upper left, right) and $r$,$i$ (lower left,right). 
\label{fig-frame}}
\end{figure}

We flux calibrated each image to the SDSS images of the same fields by 
selecting stars with magnitudes fainter than $\sim15$
to avoid saturation, and brighter than 
$\sim$20
where the $asinh$ magnitude 
used by SDSS diverges from the traditional magnitude \citep{Lupton1999}.
We used this sample to estimate the zero point magnitudes and the first order 
extinction coefficients for each 90Prime field and then applied these to the data. 
We combined the flux-calibrated images of each field to create a full exposure-time co-added image for each filter using SWarp \citep{Bertin2002}.
We cross-checked star and galaxy magnitudes once again with SDSS galaxies with $r < 21$ to verify the final quality of the calibration in the 
co-added images.
Calibrated $u$,$g$,$r$, and $i$ frames for PG0923+201 are shown in Fig.~\ref{fig-frame}.

HS0624+6907 does not overlap with the SDSS footprint, so to calibrate the frames of this field, we use the frames taken in each filter on photometric
nights as references for the frames taken on non-photometric nights.
The zero point magnitude and first order extinction coefficient is calculated from the SDSS fields taken on the same photometric nights, and so 
this photometry is also tied to the SDSS system as the rest of the sample.
We calculated photometric, Kron, and Petrosian magnitudes using the SExtractor package \citep{Bertin1996} on the full exposure-time 
flux-calibrated co-added
images of each filter. We found that optimal source extraction was achieved by considering each amplifier section of each image on a 
case-by-base basis for amps 1-8 and the extraction parameters used thus vary slightly over the full mosaic.  We give a brief summary of those parameter 
choices here.
Amps 1, 2, and 3, are typically low noise and free of bad columns. So we used $1.5\sigma$ for detection and analyze thresholds, DETECT\_THRESH and 
ANALYZE\_THRESH.
Since the fields are crowded, we use a balanced combination of background and deblending parameters, BACK\_SIZE, DEBLEND\_NTHRESH, and DEBLEND\_MINCONT, to 
detect as many individual objects and as close as possible to stars or foreground galaxies but to avoid over-resolving.
Amp 4 has noise patterns which are enhanced by resampling. To overcome this problem 
we used a moderately higher detection threshold, $\sim$3, and larger area to estimate the background with deblending 
parameters set to identify the strips as artificial bright sources. In some extreme cases we mask the noise pattern manually.
Amps 5 and 6 have a higher noise level 
than amps 1 and 2 but typically free of fixed patterns, so we 
use the same technique as amp 1 but with a higher detection threshold, $\sim$1.7.
Amps 7 and 8 behave generally like amps 5 and 6. 
We dealt with a few cases of noise patterns  using 
the same technique as amp 4, with detection a threshold of $\sim$2.3.

\begin{figure}
\epsscale{0.7}
\plotone{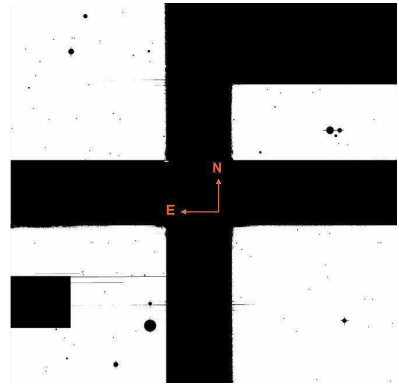}
\caption{
Example of a frame mask, for the PG0923+201 in u filter, shown in Fig.~\ref{fig-frame}, including inter-chip gaps, trap in amps 5 and 6, bad 
pixel/columns, and bright stars.
\label{fig-mask}}
\end{figure}

To remove bad columns and bright stars and their associated diffraction patterns, we created image masks which consist of:
(1) a manually generated mask for the interchip gaps; (2) a mask for a consistent trap present 
in chip 3, overlapping amps 5 and 6;
(3) a bright star mask created manually by identifying the bright stars in a given frame from the 
SExtractor output and growing the masked regions using the measured FWHM; 
(4) a bad pixel mask created using the IRAF task {\tt objmasks} 
to mask pixels with pixel values $15\sigma$ below the background; and finally,
(5) a bleed trail mask created by using the segmentation map created by 
SExtractor for a detection threshold $\sim$80-100 and very low debelending 
parameter and large background estimates, thus including all the bleed patterns. 
We created masks for each image of each field in every filter individually.
The downside of this technique is that step (5) can include some 
bright foreground galaxies into the mask, so we recover these from careful inspection of
the images in all four filters.
As an example, our mask frame for PG0923+201 in u filter 
(Fig.~\ref{fig-frame}) is shown in Fig.~\ref{fig-mask}.
Due to the large dithering of individual frames before co-adding, some regions have lower 
co-added exposure time than the full stacked image. To decrease
the biases on galaxy counting when using a certain detection limit on magnitude, 
we cut the corners and more from inner-gaps to include only the regions
with uniform co-added exposure time in each final image.
We also mask objects with centers within 8 pixels of a masked region or an edge from the final 
catalogs.
Finally we filter our catalog sources based on the internal SExtractor flags.
Here, we present the catalog of objects detected at 5$\sigma$ 
or greater in $r$ and at 3$\sigma$ or greater in $u$, $g$ and $i$ in Table~\ref{tab-catalog} as a machine readable table.  The magnitudes 
listed are the SExtractor
MAG\_BEST values, where the Kron\_fact and min\_radius parameters governing MAG\_AUTO were set to 2.5 and 1.5, respectively.
Magnitudes of 99 in this table denote non-detections in the $u$, $g$, or $i$ filter, and values of -99 label objects not observed in 
one of those filters, primarily due to masking. In the former case, the error bar gives the 1$\sigma$ detection limit for later use
with photometric redshift codes. 
For the analyses discussed in this paper, we use only galaxies detected at $5\sigma$ significance or greater.

\clearpage
\begin{turnpage}
\begin{deluxetable}{rrrrrrrrrrrrrrrrrrrr}
\tabletypesize{\scriptsize}
\tablecaption{90Prime Catalog of Galaxies in Quasar Fields
\label{tab-catalog}}
\tablewidth{0pt}
\tablehead{
\colhead{QSO field} &\colhead{RA} &\colhead{Dec} &\colhead{$u$\tablenotemark{1}} &\colhead{$\sigma_u$} &\colhead{$g$} &\colhead{$\sigma_g$} &\colhead{$r$} 
&\colhead{$\sigma_r$} &\colhead{$i$} 
&\colhead{$\sigma_i$}  &\colhead{A\tablenotemark{2}} &\colhead{$\sigma_{\rm A}$} &\colhead{B\tablenotemark{2}} &\colhead{$\sigma_{\rm B}$} 
&\colhead{radius\tablenotemark{3}} &\colhead{$\theta$\tablenotemark{3}} &\colhead{FWHM\tablenotemark{3}} 
&\colhead{class\tablenotemark{3}}
}
\startdata
HS0624p690 &98.0779 &68.8485 &23.10 &0.14 &99.00 &25.52 &21.20 &0.08 &20.65 &0.11 &1.443 &0.109 &0.840 &0.064 &2.56 &-12.91 &5.05 &0.02 \\
HS0624p690 &98.1246 &68.8445 &99.00 &27.74 &99.00 &25.52 &22.67 &0.18 &99.00 &25.54 &0.677 &0.143 &0.519 &0.110 &7.79 &-75.60 &2.93 &0.00 \\
HS0624p690 &97.2567 &68.8409 &99.00 &27.74 &99.00 &25.52 &21.93 &0.11 &21.28 &0.16 &0.852 &0.121 &0.531 &0.075 &5.67 &-85.21 &2.72 &0.61 \\
HS0624p690 &98.0474 &68.8475 &99.00 &27.74 &99.00 &25.52 &20.80 &0.06 &19.90 &0.08 &1.213 &0.080 &0.863 &0.057 &3.97 &21.11 &2.96 &0.96 \\
HS0624p690 &97.5498 &68.8449 &99.00 &27.74 &99.00 &25.52 &21.03 &0.07 &99.00 &25.54 &0.964 &0.069 &0.920 &0.065 &5.46 &-35.21 &2.37 &0.26 \\
HS0624p690 &97.6265 &68.8442 &99.00 &27.74 &99.00 &25.52 &22.11 &0.13 &22.12 &0.23 &0.799 &0.116 &0.603 &0.088 &5.97 &61.78 &2.45 &0.57 \\
HS0624p690 &97.4941 &68.8537 &99.00 &27.74 &99.00 &25.52 &21.78 &0.11 &21.71 &0.19 &0.981 &0.118 &0.682 &0.082 &5.02 &-82.06 &3.82 &0.12 \\
HS0624p690 &98.1444 &68.8458 &99.00 &27.74 &99.00 &25.52 &21.85 &0.11 &20.80 &0.12 &0.799 &0.107 &0.562 &0.075 &5.90 &73.74 &2.28 &0.73 \\
HS0624p690 &97.5849 &68.8437 &99.00 &27.74 &99.00 &25.52 &22.75 &0.17 &22.25 &0.24 &0.535 &0.129 &0.432 &0.104 &6.39 &59.89 &2.28 &0.50 \\
HS0624p690 &97.7454 &68.8461 &99.00 &27.74 &99.00 &25.52 &21.79 &0.11 &20.74 &0.12 &0.858 &0.109 &0.740 &0.094 &5.77 &29.80 &2.87 &0.10 \\
HS0624p690 &97.2771 &68.8423 &99.00 &27.74 &99.00 &25.52 &22.81 &0.18 &99.00 &25.54 &0.466 &0.144 &0.402 &0.125 &8.71 &71.16 &2.58 &0.34 \\
HS0624p690 &97.8299 &68.8482 &99.00 &27.74 &99.00 &25.52 &21.17 &0.08 &20.56 &0.11 &1.254 &0.098 &0.846 &0.066 &3.67 &46.78 &3.74 &0.21 \\
HS0624p690 &97.5091 &68.8468 &99.00 &27.74 &21.62 &0.10 &21.40 &0.09 &21.00 &0.14 &0.942 &0.086 &0.733 &0.067 &4.76 &-29.61 &2.48 &0.86 \\
HS0624p690 &97.5560 &68.8461 &99.00 &27.74 &99.00 &25.52 &22.41 &0.15 &99.00 &25.54 &0.642 &0.111 &0.559 &0.097 &6.31 &18.58 &2.16 &0.64 \\
HS0624p690 &98.0133 &68.8473 &99.00 &27.74 &99.00 &25.52 &22.77 &0.18 &21.12 &0.14 &0.753 &0.166 &0.462 &0.102 &5.87 &-88.26 &3.53 &0.23 \\
HS0624p690 &97.3695 &68.8452 &99.00 &27.74 &99.00 &25.52 &22.90 &0.18 &22.11 &0.23 &0.704 &0.156 &0.423 &0.094 &5.34 &75.93 &3.26 &0.52 \\
HS0624p690 &97.3791 &68.8490 &99.00 &27.74 &99.00 &25.52 &20.78 &0.06 &99.00 &25.54 &1.245 &0.079 &0.874 &0.055 &5.03 &28.43 &2.92 &0.20 \\
HS0624p690 &97.9586 &68.8496 &99.00 &27.74 &99.00 &25.52 &22.51 &0.16 &21.22 &0.15 &0.571 &0.125 &0.490 &0.107 &6.94 &75.39 &2.82 &0.49 \\
HS0624p690 &97.5345 &68.8568 &19.66 &0.03 &18.32 &0.02 &17.46 &0.01 &17.10 &0.02 &2.327 &0.029 &2.117 &0.027 &3.37 &-74.43 &3.91 &0.03 \\
HS0624p690 &97.5649 &68.8494 &99.00 &27.74 &99.00 &25.52 &22.12 &0.13 &21.59 &0.18 &0.542 &0.155 &0.378 &0.108 &8.39 &-13.78 &2.36 &0.36 \\
HS0624p690 &97.0202 &68.8484 &99.00 &27.74 &99.00 &25.52 &20.98 &0.07 &20.17 &0.09 &1.610 &0.121 &0.900 &0.068 &4.66 &54.87 &4.81 &0.39 \\
HS0624p690 &97.8307 &68.8524 &99.00 &27.74 &99.00 &25.52 &21.14 &0.08 &20.04 &0.09 &0.858 &0.065 &0.760 &0.057 &4.07 &79.73 &2.00 &0.96 \\
HS0624p690 &97.3351 &68.8497 &22.32 &0.09 &99.00 &25.52 &21.71 &0.10 &21.37 &0.16 &0.953 &0.101 &0.758 &0.080 &3.94 &65.93 &3.44 &0.08 \\
HS0624p690 &98.1068 &68.8543 &99.00 &27.74 &20.92 &0.07 &19.92 &0.04 &18.59 &0.04 &1.174 &0.047 &0.915 &0.037 &3.56 &80.00 &2.86 &0.94 \\
HS0624p690 &97.1526 &68.8478 &99.00 &27.74 &99.00 &25.52 &22.41 &0.15 &99.00 &25.54 &0.676 &0.147 &0.501 &0.109 &7.64 &-77.44 &3.48 &0.47 \\
HS0624p690 &97.9007 &68.8552 &99.00 &27.74 &99.00 &25.52 &20.11 &0.05 &19.11 &0.06 &1.381 &0.062 &0.896 &0.040 &3.47 &9.03 &3.23 &0.94 \\
HS0624p690 &97.7168 &68.8546 &99.00 &27.74 &99.00 &25.52 &20.83 &0.07 &20.13 &0.09 &1.116 &0.072 &0.897 &0.058 &4.24 &-75.39 &2.69 &0.53 \\
HS0624p690 &97.8514 &68.8552 &99.00 &27.74 &99.00 &25.52 &21.59 &0.10 &21.05 &0.14 &0.912 &0.125 &0.780 &0.107 &6.80 &0.23 &4.54 &0.00 \\
HS0624p690 &97.9116 &68.8545 &99.00 &27.74 &99.00 &25.52 &22.63 &0.16 &21.52 &0.17 &0.705 &0.127 &0.522 &0.094 &4.74 &74.68 &3.43 &0.64 \\
HS0624p690 &98.0572 &68.8560 &22.87 &0.13 &99.00 &25.52 &21.29 &0.09 &20.81 &0.12 &1.592 &0.142 &0.783 &0.070 &4.98 &87.82 &5.97 &0.00 \\
HS0624p690 &97.6051 &68.8607 &99.00 &27.74 &99.00 &25.52 &22.02 &0.12 &22.28 &0.24 &0.863 &0.114 &0.664 &0.088 &5.13 &85.51 &2.90 &0.12 
\enddata
\tablenotetext{1}{MAG\_BEST from SExtractor}
\tablenotetext{2}{semimajor and semiminor axes in arcsec}
\tablenotetext{3}{SExtractor parameters KRON\_RADIUS, THETA\_J2000 (in degrees), FWHM, and CLASS\_STAR as discussed in the text and in \cite{Bertin2002}}
\end{deluxetable}
\end{turnpage}
\clearpage

\begin{figure}
\epsscale{1.0}
\plotone{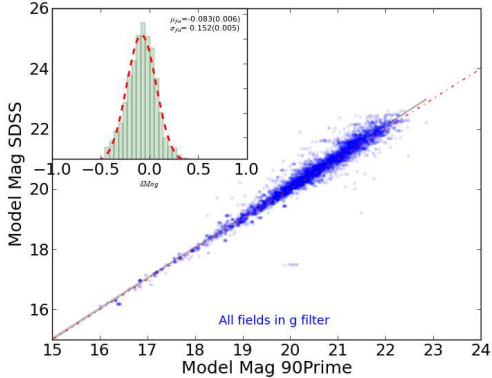}
\caption{
Example of error estimates from photometric calibration process in g filter from all fields.
\label{fig-Error-calibration}}
\end{figure}

We estimate systematic error and scatter due photometric calibration for all 
four filters using a final comparison with the SDSS photometry.  For this comparison, we 
use all objects detected in our catalogs at $\geq$5$\sigma$ and with a SExtractor classification parameter (CLASS\_STAR, described in the next section)
likely to be a galaxy, $<$0.3.
We also require the relative error in the SDSS magnitude 
to be smaller than 1\%.
The values of the systematic shifts are small, typically less than $0.04$. They are listed in Table~\ref{tab-Flux-Calibration-Error} and plotted
for the $g$ band in Fig.~\ref{fig-Error-calibration}.  Also in Table~\ref{tab-Flux-Calibration-Error} we list the 95\% confidence on the 
systematic offset, along with the total photometric statistical uncertainties in each band and their 95\% confidence interval.  
The total statistical uncertainties are obtained from adding the systematic offsets from SDSS 
in quadrature with the standard deviation of the 
distribution of SExtractor magnitude
errors for each filter. 
The mean and standard deviation of those distributions are listed separately for each field in 
Table \ref{tab-All-Flux-Calibration-Error}.
\begin{deluxetable}{lcccc}
\tablecaption{Flux Calibration Errors
\label{tab-Flux-Calibration-Error}}
\tablewidth{0pt}
\tablehead{
\colhead{filter}
&\colhead{Systematic err.}
&\colhead{95\% conf.}
&\colhead{Total}
&\colhead{95\% conf.}
}
\startdata
$u$ & -0.039 & 0.011 & 0.305 & 0.009 \\
$g$ & -0.083 & 0.006 & 0.152 & 0.005 \\
$r$ & -0.028 & 0.008 & 0.122 & 0.007 \\
$i$ & -0.022 & 0.013 & 0.066 & 0.006 \\
\enddata
\end{deluxetable}

\begin{deluxetable}{lccccc}
\tablecaption{Individual Flux Calibration Errors
\label{tab-All-Flux-Calibration-Error}}
\tablewidth{0pt}
\tablehead{
\colhead{field}
&\colhead{filter}
&\colhead{Systematic err.}
&\colhead{95\% conf.}
&\colhead{Total}
&\colhead{95\% conf.}
}
\startdata
MRK586 & $u$ & -0.061 & 0.016 & 0.328 & 0.012 \\
MRK586 & $g$ & -0.026 & 0.012 & 0.253 & 0.010 \\
MRK586 & $r$ & 0.007 & 0.016 & 0.139 & 0.014 \\

PG0844+349 & $u$ & 0.009 & 0.013 & 0.293 & 0.010 \\
PG0844+349 & $r$ & -0.031 & 0.008 & 0.131 & 0.007 \\

PG0923+201 & $u$ & 0.003 & 0.013 & 0.356 & 0.011 \\
PG0923+201 & $g$ & -0.062 & 0.012 & 0.169 & 0.011 \\
PG0923+201 & $r$ & 0.022 & 0.012 & 0.144 & 0.010 \\
PG0923+201 & $i$ & 0.119 & 0.020 & 0.199 & 0.016 \\

PG0953+414 & $u$ & 0.005 & 0.011 & 0.286 & 0.009 \\
PG0953+414 & $g$ & -0.030 & 0.008 & 0.134 & 0.006 \\
PG0953+414 & $r$ & -0.008 & 0.010 & 0.114 & 0.009 \\
PG0953+414 & $i$ & 0.062 & 0.013 & 0.131 & 0.010 \\

PG1116+215 & $u$ & -0.019 & 0.013 & 0.317 & 0.010 \\
PG1116+215 & $g$ & -0.029 & 0.025 & 0.118 & 0.015 \\
PG1116+215 & $r$ & -0.019 & 0.041 & 0.070 & 0.019 \\
PG1116+215 & $i$ & 0.095 & 0.018 & 0.171 & 0.014 \\

PG1307+085 & $u$ & 0.034 & 0.010 & 0.292 & 0.008 \\
PG1307+085 & $g$ & -0.046 & 0.010 & 0.148 & 0.008 \\
PG1307+085 & $r$ & -0.014 & 0.006 & 0.071 & 0.005 \\
PG1307+085 & $i$ & 0.067 & 0.018 & 0.145 & 0.012 \\

PG1404+226 & $u$ & -0.117 & 0.016 & 0.376 & 0.013 \\
PG1404+226 & $g$ & -0.076 & 0.010 & 0.172 & 0.008 \\
PG1404+226 & $r$ & -0.032 & 0.008 & 0.134 & 0.007 \\
PG1404+226 & $i$ & 0.064 & 0.022 & 0.171 & 0.018 \\

PG1444+407 & $u$ & -0.083 & 0.014 & 0.284 & 0.012 \\
PG1444+407 & $g$ & -0.089 & 0.008 & 0.148 & 0.007 \\
PG1444+407 & $r$ & -0.036 & 0.018 & 0.111 & 0.016 \\
PG1444+407 & $i$ & 0.006 & 0.013 & 0.145 & 0.011 \\

PG1545+210 & $u$ & -0.005 & 0.007 & 0.331 & 0.006 \\
PG1545+210 & $g$ & -0.091 & 0.014 & 0.165 & 0.013 \\
PG1545+210 & $r$ & -0.017 & 0.011 & 0.093 & 0.008 \\
PG1545+210 & $i$ & 0.036 & 0.017 & 0.161 & 0.014 \\

PG1612+261 & $u$ & -0.088 & 0.014 & 0.305 & 0.011 \\
PG1612+261 & $g$ & -0.110 & 0.010 & 0.128 & 0.008 \\
PG1612+261 & $r$ & -0.045 & 0.011 & 0.108 & 0.008 \\
PG1612+261 & $i$ & 0.033 & 0.012 & 0.185 & 0.010 \\

Q2141+175 & $g$ & -0.093 & 0.010 & 0.155 & 0.007 \\
Q2141+175 & $r$ & -0.058 & 0.012 & 0.100 & 0.011 \\
Q2141+175 & $i$ & 0.010 & 0.021 & 0.185 & 0.014 \\
\enddata
\end{deluxetable}

\subsubsection{Star-Galaxy Separation and Filtering}
\begin{figure}
\epsscale{1.0}
\plotone{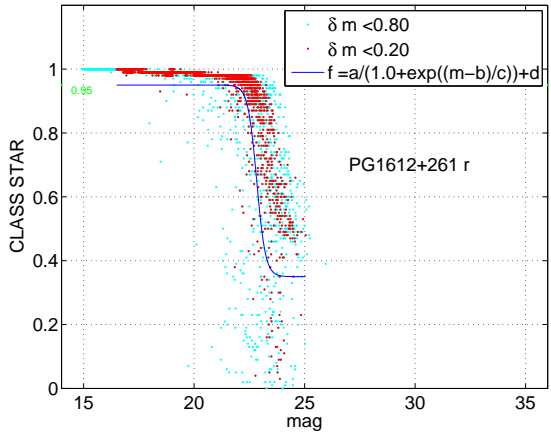}
\caption{
CLASS\_STAR versus r for PG1612+261 with the fit to Equ.~\ref{equ-class}.
Cyan(red) points are points for which the difference between
the input and output stellar magnitudes in the Monte-Carlo simulations
was less than 0.8(0.2).  The fit is designed to find the best inner envelope 
for the red data reaching asymptotic value of $0.95$ at lowest magnitude. 
\label{fig-classstar}}
\end{figure}

SExtractor calculates a CLASS\_STAR flag based on core geometry of the light distribution of the detected objects, where
values near 1 indicate a centrally concentrated star and those near 0 indicate a galaxy.  This parameter is sensitive to the
value of the input parameter SEEING\_FWHM.  We measure this input parameter from objects clearly classified as stars 
($\rm{CLASS\_STAR} > 0.9$)
in initial photometric catalogs,
which were created from a first pass with SExtractor on the masked images. 
To improve the accuracy of the CLASS\_STAR star-galaxy separation output parameter, we performed this measurement iteratively until 
we reached convergence between the input value of the SEEING\_FWHM and the measured FWHM of the stellar catalog sources. 
We achieved convergence of $<$0.1\arcsec\ in 1-2 iterations.
In order to 
determine how to use this CLASS\_STAR parameter to best distinguish stars from galaxies in our final catalogs, we 
performed Monte-Carlo simulations of our data set using synthetic input stars that we then extract from the data frames using the 
same SExtractor parameters as we used on the data.
Using the IRAF task {\tt mkobjects}, we generated synthetic stars with magnitudes between 15 and 27 and with PSFs that matched those of stellar 
objects in the 
post-iteration catalogs.
We randomly distributed these synthetic stars on chips 1 and 4 (amps 1,2,7, and 8) of each image, to bracket the best and worst quality CCDs in the mosaic.
After extracting these simulated stars, we see the typical behavior of the CLASS\_STAR parameter, i.\ e. that it is a reliable star-galaxy separator at
bright magnitudes but at a particular threshold magnitude, CLASS\_STAR is spread evenly between 1.0 and some saturation value, 0.35 for our 90Prime frames.
Thus, at magnitudes dimmer than the threshold, we cannot rely on this parameter alone to separate galaxies from stars.  This threshold magnitude
varies across our data frames, but for each frame, we fit a function to the lower envelope of CLASS\_STAR versus magnitude, $m$:
\begin{equation}
{\rm CLASS\_STAR} = a/(1.0+ \exp((m-b)/c) ) + d
\label{equ-class}
\end{equation}
where $m$ is the magnitude in the relevant filter, and $a$, $b$, $c$, and $d$ are fitted parameters. 
We use the fits to this expression in our
final star-galaxy separation algorithm.
Figure~\ref{fig-classstar} shows an example of this fitting.

\begin{figure}
\epsscale{1.0}
\plotone{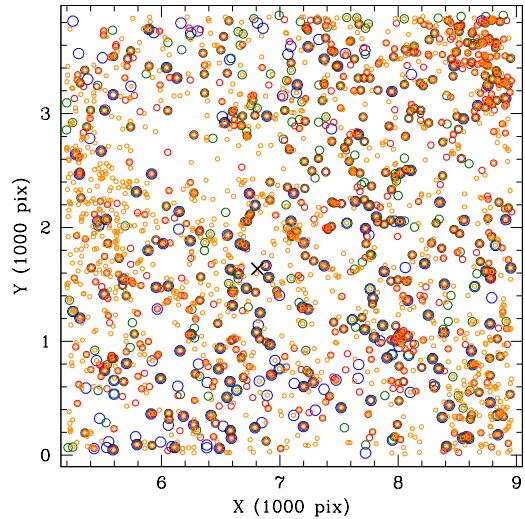}
\caption{Blue, green, red, and orange circles show the positions of galaxies detected in $u$, $g$, $r$, and $i$
at $>5\sigma$ significance with magnitude brighter than the limiting magnitude
listed in column 3 of Table~\ref{tab-mlim} for chip 1 of the PG1444+407 field.
Black cross marks the position of PG1444+407 itself.
\label{fig-pg1444pos}}
\end{figure}

Equ.~\ref{equ-class} is used to define a cut in the CLASS\_STAR parameter as function of source magnitude.  
We find that using this function as a strict cut results in many faint sources lost from our catalogs,  so we relax this
criterion and rely on
a color based criterion  as well.
To construct the color criterion, we use the unambiguous stellar sources in the matched catalogs, those with CLASS\_STAR $>$ 0.95
in all filters, along with the sources in the Gunn-Stryker Atlas \citep{Gunn1983} to define
the stellar locus in color-color space.  We then exclude from our catalogs any source that lies within a specified 
distance from that locus in both $g-r$ versus $u-g$ and $r-i$ versus $g-r$.  
This distance is allowed to vary in order to give the best match in number counts to our 
adopted luminosity functions as described below, but it is typically 0.1-0.3.
In order to use the color information for sources with only one available color, we use a technique similar to that described by \cite{Coil2004}. 
The binned distributions of the stars in our catalogs in each color,
normalized to have a maximum value of unity, are used to assign a probability of being a star, $p_{s}$.
Objects having a $p_{s}$ above a threshold value are excluded from the final galaxy catalogs. This threshold is also allowed to vary slightly, 
typically between 0.7 and 0.8, to get
good agreement in the number counts in each filter.
The resulting number counts from these final catalogs, shown below, give us confidence in this star-galaxy separation 
technique. In Figure~\ref{fig-pg1444pos}, we show the positions of the 5$\sigma$ galaxies in $u$, $g$, $r$, and $i$ in the portion of the field of
PG1444+407 covered by chip 1, along with the position of the quasar itself.

\subsubsection{Limiting Magnitudes}
\begin{figure}
\epsscale{1.0}
\plotone{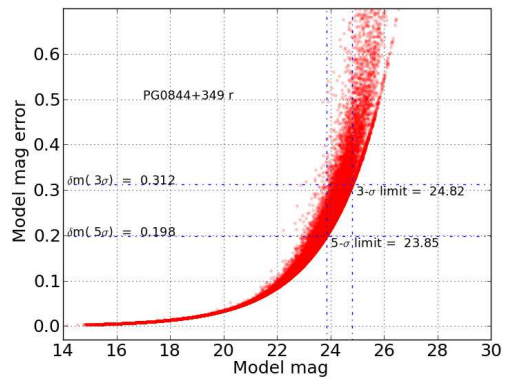}
\caption{Example of method (1) for calculating limiting magnitudes at various significance levels. 
\label{fig-mlim1}}
\end{figure}

\begin{figure}
\epsscale{1.0}
\plotone{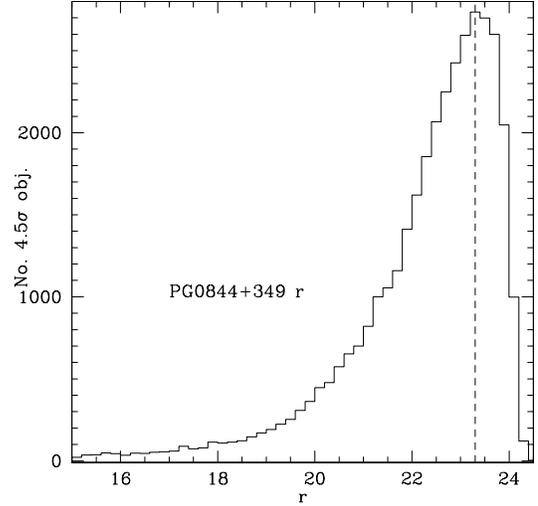}
\caption{Example of method (3) for calculating completeness using the magnitude distribution of all catalog objects detected at $>4.5\sigma$.
\label{fig-mlim3}}
\end{figure}

We characterize the depth of our data frames in two ways: (1)
by finding the magnitude at which the lower envelope of the 
distribution of magnitude errors for stars in the catalogs equals 0.217, or 5$\sigma$
and (2) by using the magnitude distribution of all catalog objects detected at
$>4.5 \sigma$.
We show examples of each in Figures~\ref{fig-mlim1} and \ref{fig-mlim3} 
and tabulate these values for each co-added frame in Table~\ref{tab-mlim}.
Because we used slightly modified SExtractor parameters for the different amps of the full mosaic, we calculated the limiting magnitudes
separately for each amp using method (1).  We find that the values for all other amps agree with that of amp 1 to within $<0.03$ magnitudes on average,
so we report only values for that amplifier.
\begin{deluxetable}{lcccc}
\tablecaption{Limiting Magnitudes
\label{tab-mlim}}
\tablewidth{0pt}
\tablehead{
\colhead{QSO}  &\colhead{filter}  &\colhead{5$\sigma$ m$_{\rm lim}$}  &\colhead{3$\sigma$ m$_{\rm lim}$}
   &\colhead{m(complete)} \\
\colhead{} &\colhead{}  &\colhead{method (1)}  &\colhead{method(1)}
 &\colhead{method (2)}
}
\startdata
HS0624+6907 &$u$ &23.8 &23.9 &22.9\\
HS0624+6907 &$g$ &21.9 &21.9 &20.7\\
HS0624+6907 &$r$ &23.1 &23.4 &22.1\\
HS0624+6907 &$i$ &22.0 &23.1 &21.7\\
MRK586 &$u$ &22.9 &24.0 &22.5\\
MRK586 &$g$ &23.6 &24.5 &22.9\\
MRK586 &$r$ &23.3 &24.4 &23.1\\
PG0844+349 &$u$ &23.9 &24.5 &23.3\\
PG0844+349 &$r$ &24.0 &25.1 &23.3\\
PG0923+201 &$u$ &23.7 &24.8 &23.5\\
PG0923+201 &$g$ &23.6 &24.7 &23.3\\
PG0923+201 &$r$ &22.7 &23.8 &22.5\\
PG0923+201 &$i$ &22.8 &24.0 &22.3\\
PG0953+414 &$u$ &24.1 &25.2 &23.9\\
PG0953+414 &$g$ &22.9 &24.1 &22.9\\
PG0953+414 &$r$ &23.3 &24.4 &23.1\\
PG0953+414 &$i$ &23.6 &24.7 &22.7\\
PG1116+215 &$u$ &23.7 &24.8 &23.5\\
PG1116+215 &$g$ &22.0 &23.0 &21.7\\
PG1116+215 &$r$ &22.2 &23.3 &22.1\\
PG1116+215 &$i$ &22.8 &23.9 &22.5\\
PG1307+085 &$u$ &23.5 &24.6 &23.3\\
PG1307+085 &$g$ &22.7 &23.9 &22.7\\
PG1307+085 &$r$ &23.1 &24.2 &23.1\\
PG1307+085 &$i$ &22.9 &24.0 &22.7\\
PG1404+226 &$u$ &23.6 &24.6 &23.3\\
PG1404+226 &$g$ &23.6 &24.8 &23.5\\
PG1404+226 &$r$ &24.0 &25.1 &23.5\\
PG1404+226 &$i$ &22.8 &23.9 &22.5\\
PG1444+407 &$u$ &23.4 &24.6 &23.3\\
PG1444+407 &$g$ &23.4 &24.6 &23.3\\
PG1444+407 &$r$ &23.2 &24.4 &23.1\\
PG1444+407 &$i$ &22.8 &24.0 &22.7\\
PG1545+210 &$u$ &23.8 &24.9 &23.5\\
PG1545+210 &$g$ &23.5 &24.7 &23.3\\
PG1545+210 &$r$ &23.1 &24.2 &22.9\\
PG1545+210 &$i$ &22.8 &23.9 &22.3\\
PG1612+261 &$u$ &23.3 &24.4 &23.1\\
PG1612+261 &$g$ &23.9 &25.0 &23.7\\
PG1612+261 &$r$ &23.6 &24.7 &23.5\\
PG1612+261 &$i$ &23.1 &24.3 &22.9\\
Q2141+175 &$g$ &23.8 &24.9 &23.7\\
Q2141+175 &$r$ &23.4 &24.5 &23.3\\
Q2141+175 &$i$ &22.6 &23.8 &22.3
\enddata
\end{deluxetable}


We treat method (1) as our estimator of the limiting magnitude at a particular level of significance.  
Because we use stars only for method (1), it is affected by our ability to recover faint extended galaxies from the images.  
Method (2) therefore best characterizes the depth of the galaxy catalogs, since those objects dominate the number counts at the faintest magnitudes, assuming
our star-galaxy separation technique is robust. We discuss this further below, in Section~\ref{sec-number}.

Results for each field/filter with each method are listed in Table~\ref{tab-mlim}.
For reference, the SDSS limiting magnitudes in $(u,g,r,i)$ are (22.0, 22.2, 22.2, 21.3), corresponding to 95\% completeness limits.
These values may be compared with column 4, 
where we list the $3\sigma$ limiting magnitudes found via method (1).  
The stacked $g$ band image of the HS0624+6907 field was created from a combination of data taken on  photometric and non-photometric nights, calibrated
as described above. The resuling image in this particular field and filter had a resulting level of noise that necessitated a high detection threshold
within SExtractor to avoid numerous spurious detections around bright sources. Thus, the 3- and 5$\sigma$  
limiting magnitudes are reported here to be the same, as there are no sources extracted at less than 4$\sigma$. This also results in a shallow field
compared to the other photometry reported here.
Note that the limiting magnitudes in the 90Prime data 
are significantly fainter than SDSS in most other fields and filters.

\section{Number Counts}
\label{sec-number}
\begin{figure}
\epsscale{1.0}
\plotone{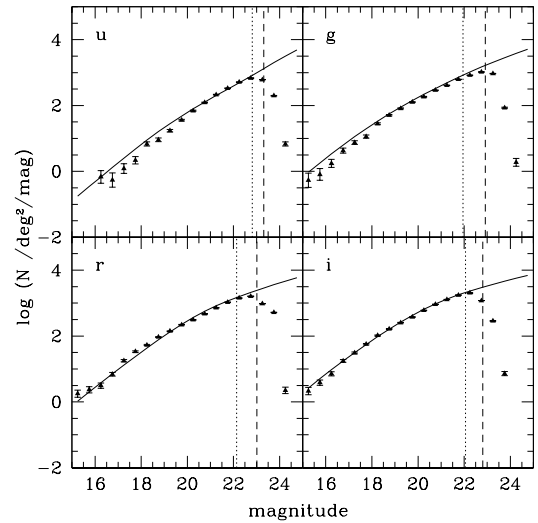}
\caption{Number of galaxies in 0.5 magnitude bins per square degree in our 90Prime fields in $u$, $g$, $r$, and $i$.
Solid lines are the number counts predicted by the luminosity function in each band.
The vertical dotted and dashed lines show the limiting magnitudes for Samples 1 and 2 respectively,
listed in Table~\ref{tab-sample} and discussed in Section~\ref{sec-corr}.
\label{fig-ncounts}}
\end{figure}

\begin{figure}
\epsscale{1.0}
\plotone{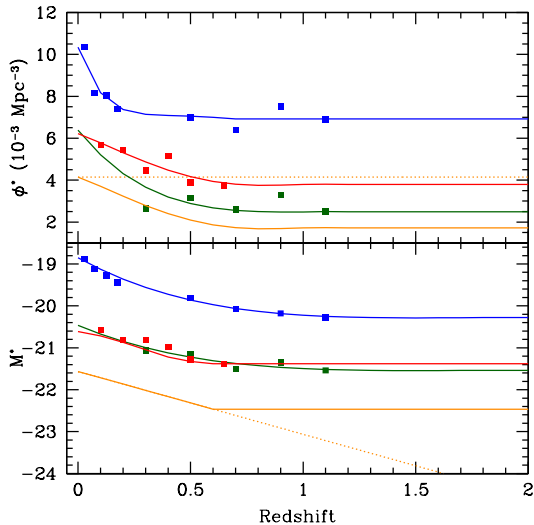}
\caption{Polynomial fits (solid curves) to luminosity function parameter values $\phi^{*}$ and $M^{*}$ 
(squares) from
\cite{Prescott2009} ($u$ band, blue), \citep{Faber2007} ($B$ band, green), \cite{Cool2012} ($r$ band, red), and
For $i$ band (orange) the \cite{Loveday2012} parametrized model for $M^{*}$ is set at the $z=0.6$ value
for larger redshifts, and for $\phi^{*}$ it follows the redshift evolution of the $r$ band from its $z=0$ value.
Dashed curves show \cite{Loveday2012} parametrized models for the redshift evolution of each parameter in $i$ 
extrapolated to $z>0.5$.
\label{fig-lf}}
\end{figure}

In Figure~\ref{fig-ncounts} we show the galaxy number counts in all four bins from our 90Prime fields, over
a total of 7.35, 7.45, 8.95, and 7.03 deg$^2$ in $u$, $g$, $r$, and $i$, respectively.  
The solid lines in Figure~\ref{fig-ncounts} show the predicted number of galaxies for our adopted
luminosity functions in each band.
Luminosity function parameters adopted are shown in Fig.~\ref{fig-lf}.
The curves shown here are polynomial fits to the parameter values derived as a function of redshift from several surveys:
in $u$, we adopt the parameters found by \cite{Prescott2009} from the DEEP2 and SDSS $u$-band Galaxy Survey;
and for the $r$  band, we adopt the parameters found by \cite{Cool2012} from the AGES survey.
For the $g$  band, we adopt the $B$ band values from the DEEP2 survey \citep{Faber2007}, using the $g$ band luminosity function parameter
measurements of \cite{Loveday2012} to constrain our fits at $z=0$. We make this choice because:
(a) the wavelength coverage of $B$ overlaps that of the SDSS $g$ filter; (b) the agreement between
these B band luminosity function parameters and the directly measured parameters in $g$ from \cite{Loveday2012} is good at $z<0.5$;
and (c) while no AGN with
$z>0.5$ are considered in this work, the \cite{Faber2007} study provides a 
more robust and convenient way to parametrize the luminosity function to redshifts beyond $z=0.5$ than an 
extrapolation of the \cite{Loveday2012} measurements, enabling us to estimate
the numbers of background galaxies reliably.
For the $i$ band, we adopt a modification of the evolution derived by \cite{Loveday2012}.  
The $M^{*}$ follows the parametric solution found by these authors, but we fix its
value at $z>0.6$ to the $z=0.6$ value.  For $\phi^{*}$, we find good agreement with our
galaxy number counts by using the \cite{Loveday2012} value at $z=0$ and adopting the redshift evolution
for the $r$ band, noting that the \cite{Loveday2012} parametrized solution found no redshift evolution in this band. 
The $M^{*}$ and $\phi^{*}$ evolution in $i$ found by \cite{Loveday2012} and extrapolated to 
$z> 0.5$ is shown in Fig.~\ref{fig-lf}.
For $\alpha$, the values from these various studies are (-1.0, -1.3, -1.05, -1.12) for $(u,B,r,i)$.
We extrapolate the highest redshift values from each study to all higher redshifts for our calculations, although
at the depths of our survey, we are generally not sensitive to galaxies with $z\gtrsim 1.5$.

\section{Correlation Functions}
\label{sec-corr}
To examine the galaxy autocorrelation function and the  galaxy-quasar cross-correlation signal in the 90Prime fields, we
use a chip-by-chip approach to the calculations.  Due to the large interchip gaps in the 90Prime mosaic, the masked region 
in chip 3, and the often masked amplifier 4 on chip 2 (see Fig.~\ref{fig-mask}), the correlation functions described below
contain significant structure when calculated at scales that span these regions, 
even with a large simulated set of random points.  
To verify this, we performed a series of simulations with 5000 random data points created with regions identical to those in the real data. 
With the random data, we expect zero correlation but find that there is significant structure in the correlation function when the masks are applied, and so we 
calculate the correlation functions in the real data on scales that minimize the effect of the masks introducing structure into the 
correlation functions that is unrelated to the galaxy distributions.

To perform the correlation function calculations to a uniform depth, we confine galaxy magnitudes to the limit defined by
the shallowest field in the sample. These depths are listed in Table~\ref{tab-sample} for the two cases we consider:
Sample 1 is the sample of all quasar fields with the HS0624+6901 
$g$ band excluded due to its shallow depth ($g=21.79$) and Sample 2 is a sample of deeper fields, constructed by leaving out
other shallow fields in each filter to achieve the depth listed in Table~\ref{tab-sample}.
The Sample 1 and 2 limiting magnitudes are also shown as vertical dotted and dashed lines in Figure~\ref{fig-ncounts}.
In our calculations, we will consider other AGN in the primary sample fields with redshifts up to $z=0.5$. 
Thus, in Table~\ref{tab-sample} we also list the 
absolute magnitudes relative to M$^{*}$ reached for $z=0.5$ galaxies in Samples 1 and 2 in order to demonstrate the coverage of
the galaxy luminosity function at the redshifts of the AGN in the overall sample.
\begin{deluxetable}{llc}
\tablecaption{Sample Limiting Magnitudes
\label{tab-sample}}
\tablewidth{0pt}
\tablehead{
\colhead{Filter} &\colhead{m$_{\rm lim}$} &\colhead{ M$_{\rm lim}$ - M$^{*}$\tablenotemark{1} }
}
\startdata
\multicolumn{3}{c}{Sample 1} \\
\hline
$u$   &22.82 &-1.1\\
$g$   &21.95 &-0.8\\
$r$   &22.14 &0.3\\
$i$   &22.06 &1.6\\
\hline
\multicolumn{3}{c}{Sample 2}\\
\hline
$u$   &23.31 &-0.6\\
$g$   &22.92 &0.2\\
$r$   &23.01 &1.1\\
$i$   &22.81 &2.3\\   
\enddata
\tablenotetext{1}{For $z=0.5$}
\end{deluxetable}

\subsection{Galaxy-Galaxy Clustering}
We calculate the angular correlation function of the
galaxy-galaxy sample,
\begin{equation}
\omega (\theta)_{\rm true} = A_{\rm gg} \theta^{(1-\gamma)}
\label{equ-ang}
\end{equation}
using the estimator of \cite{LandySzalay1993}:
\begin{equation}
\omega (\theta)_{\rm obs} = \frac{N_{\rm gg}(\theta) - 2N_{\rm gr}(\theta) - N_{\rm rr}(\theta)}{N_{\rm rr}(\theta)}
\end{equation}
where $N_{\rm gg}$ and $N_{\rm rr}$ are the pairwise distances between
all galaxies in the sample
and all random positions in the survey area at a given angular separation, $\theta$, 
normalized by the total numbers of galaxy-galaxy and random-random pairwise distances.
The $N_{\rm gr}$ term is the normalized number of
pairwise distances between all galaxies and random points in each $\theta$ bin.

The random points are generated by Monte Carlo simulations which place galaxies at
random positions within the areas covered by the 90Prime images.  These catalogs of
random positions are masked in the same way as the galaxy catalogs.  We generated 10 realizations of
each field in each filter, with  $\sim$40,000 points in each realization.
The Poisson error in each bin is given by
\begin{equation}
\Delta \omega (\theta) = \frac{ 1 + \omega(\theta)}{\sqrt{N_{\rm gg}(\theta)}}.
\end{equation}
This estimator of the angular correlation function must be corrected for
the integral constraint, which requires the function to reach zero at the survey edges, so  that
\begin{equation}
\omega (\theta)_{\rm true} = \omega (\theta)_{\rm obs} + C
\end{equation}
where
\begin{equation}
C = \frac{1}{\Omega^2} \int \int \omega(\theta) d\Omega d\Omega = \frac{\sum N_{\rm rr}(\theta) A_{\rm gg} \theta^{(1-\gamma)}}{\sum N_{\rm rr} (\theta)} \label{equ-IC} 
\end{equation}
\citep{GrothPeebles1977,Infante1994}.

In each filter, we solve for A$_{\rm gg}$ and $\gamma$ independently using a least squares technique and
derive the uncertainties in the parameters by finding the values of each that yield $\chi^{2}_{\rm min} + 1$.

\subsection{Galaxy-Quasar Clustering}
\label{sec-omegagq}
For galaxy-quasar clustering, the autocorrelation function does not apply, so we follow \cite{CroomShanks1999} and
\cite{Brown2001}
\begin{equation}
\omega (\theta)_{\rm true} = \omega (\theta)_{\rm obs} + C = \frac{N_{\rm gq}(\theta)}{N_{\rm qr}(\theta)} - 1  + C  
\label{equ-omegagq}
\end{equation}
where $N_{\rm gq}$ and $N_{\rm qr}$ are the numbers of galaxy-quasar and quasar-random pairs at each value of $\theta$, normalized
by the  numbers of galaxies and random points respectively and
$C$ is defined as in Equ.~\ref{equ-IC}.

Here we solve for \agq\ for a fixed value of $\gamma$ 
and again derive the uncertainties in the parameters by finding the values of each that yield $\chi^{2}_{\rm min} + 1$.

\subsubsection{Correlation Length}
The galaxy-galaxy and the galaxy-quasar angular correlation functions 
can be used to derive the correlation length, $r_{0}$, of the spatial correlation function
\begin{equation}
\xi = \left( \frac{r}{r_0} \right)^{\gamma} (1+z)^{-(3+\epsilon)} \label{equ-xi}
\end{equation}
for galaxy-galaxy and galaxy-quasar clustering
using Limber's equation \citep{Limber1953}, which relates $r_{0}$ to the amplitude of the angular correlation function given in 
Equation~\ref{equ-ang}:
\begin{equation}
A = k  r_0^{\gamma} \frac{ \int_{0}^{\infty} F(z) D_{\theta}^{(1-\gamma)}(z) N_x(z)N_g(z)g(z) dz}{ \int_{0}^{\infty}  N_x(z)N_g(z) dz} 
\label{equ-limber}
\end{equation}
where
\begin{equation}
k = \sqrt(\pi) \frac{ \Gamma \left( \frac{\gamma-1}{2} \right) } { \Gamma \left( \frac{\gamma}{2} \right) } 
\end{equation}
and $F(z)$ describes the redshift evolution of the correlation function in Equ.~\ref{equ-xi} with 
$\epsilon=0$ or $\gamma-3$ for clustering fixed in physical or comoving coordinates, or
$\epsilon=\gamma-1$ for clustering according to linear theory.  Finally,
\begin{equation}
g(z) = \frac{H_0}{c} [ (1+z)^3 ( 1 + \Omega_m z + \Omega_{\Lambda} ( (1+z)^{-2} -1)^{\frac{1}{2}} ]
\end{equation}
and $N_x$ corresponds to either $N_g$ or $N_q$ depending on whether we are considering
galaxy-galaxy or galaxy-quasar clustering.

\begin{figure}
\epsscale{1.0}
\plotone{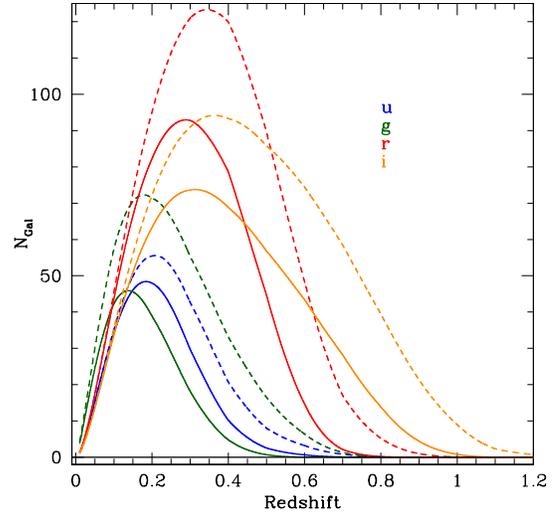}
\caption{Number of galaxies in $u$, $g$, $r$, $i$ frames in all 90Prime fields from integration of a Schechter luminosity
function with parameters as shown in Figure~\ref{fig-lf}, integrated to the limiting magnitudes listed in
column 3 of Table~\ref{tab-mlim}.  Solid lines show Sample 1 and dashed lines represent Sample 2.
\label{fig-zdist}}
\end{figure}

The redshift distribution of the galaxies in our 90Prime Samples 1 and 2, calculated from the
Schechter luminosity function parameters shown in Figure~\ref{fig-lf} and the magnitude limits
determined by the values listed in Table~\ref{tab-sample} in each of the 
four bands is shown in Figure~\ref{fig-zdist}.

\subsubsection{Correlation Amplitude}
\label{sec-bgq}
We also calculate the amplitude of the galaxy-quasar correlation function, \bgq\ using the technique described by
\cite{Ellingson1991}, \cite{Finn2001}, \cite{McLure2001}, and recently by 
\cite{RamosAlmeida2013}.
We first begin with the angular correlation function, which describes the 
clustering of galaxies about some particular object using the following relation:
\begin{equation}
n(\theta) d\Omega = N_{g}[1+w(\theta)] d\Omega. 
\end{equation}
Here, $n(\theta)d\Omega$ represents the number of galaxies within a solid angle d$\Omega$ at an angular 
distance $\theta$ from the object. The term $N_{g}$ represents the average galaxy surface density, 
and $w(\theta)=A_{gq}\theta^{1-\gamma}$ represents the angular cross-correlation function discussed above. 
To calculate \bgq, we first determine the amplitude, \agq,
from the integral of the angular correlation function within a fixed angular radius, $\theta$,
\begin{equation}
A_{gq}=\frac{N_{t}-N_{b}}{N_{b}}\frac{(3-\gamma)}{2}\theta^{\gamma-1},
\label{equ-agq}
\end{equation}
where $N_{t}$ represents the quasar field galaxy counts, and $N_{b}$ represents the background 
galaxy counts both within angular distance $\theta$ of the quasar.  
For each quasar field, we count the galaxies within a specified radius around the
quasar, down to the magnitude limit specified by the amplifiers on the chip, $m_{\rm lim}$.  
To construct the control fields, we use all other regions of the mosaic of the reference quasar as well
as any non-quasar regions of the images of our other quasar fields. We place grids of overlapping control
fields with the same radius as our reference quasar in each of these regions and count the galaxies
brighter than $m_{\rm lim}$. The galaxy counts in 
these regions may be affected by other galaxy clusters and large scale structures in the quasar fields.
We discuss these in detail for each quasar field in Appendix~\ref{sec-notes}.
To mitigate this problem we take a median of all the counts from these regions and use this as our value for
$N_{b}$.  

The angular covariance amplitude, \agq, quantifies any excess in galaxy counts 
within the quasar field as compared with the background counts of the control field.  
To de-project the angular 
cross-correlation function into its spatial equivalent, we use:
\begin{equation}
n(r) dV=\rho_{g}[1+\xi(r)] dV,
\end{equation}
where $n(r)$ and $\rho_{g}$ represent the total and expected average galaxy counts within volume $dV$,
respectively, and the spatial cross-correlation function, $\xi(r)$, is defined as in Equ.~\ref{equ-xi}, with
amplitude \bgq :
\begin{equation} 
\xi (r)={\rm B}_{gq}r^{-\gamma},
\end{equation}
where $\gamma=1.77$ \citep{GrothPeebles1977}.  
\bgq\ can be obtained using the following equation relating the spatial 
clustering amplitude to the angular clustering amplitude \citep{LongairSeldner1979}:
\begin{equation}
{\rm B}_{gq}=\frac{N_{g}A_{gq}}{\Phi(m_{\rm lim},z)I_{\gamma}}\left[\frac{D}{1+z}\right]^{\gamma-3}. \label{equ-bgq}
\end{equation}
Here, $D$ is the angular diameter distance to the quasar, and $\phi(z)$ is the galaxy 
luminosity function integrated to the limiting absolute 
magnitude of the data at the quasar redshift. The quantity $I_{\gamma}$ is a constant of integration
equivalent to $\frac{2^\gamma}{(\gamma-1)} \frac{\Gamma(\gamma+1/2)}{\Gamma(\gamma)}$.

The errors in \agq\ and \bgq\ can be calculated using the following equations:
\begin{equation}
\frac{\Delta A_{gq}}{A_{gq}}=\frac{\Delta B_{gq}}{B_{gq}}= \frac{1}{N_{t}-N_{b}}
\left[(N_{t}-N_{b})+1.3^{2}N_{b}\right]^\frac{1}{2}. 
\end{equation}
\citep{YeeLopezCruz1999}.
We normalize the galaxy counts using the luminosity function calculated with the redshift-dependent parameters
shown in Fig.~\ref{fig-lf}.

To calculate the predicted average galaxy background counts, $N_{g}$, 
we integrate the luminosity function using the same redshift-dependent parameters: 
\begin{equation}
N_{g}=\int_{0}^{z} \int_{-\infty}^{M_{\rm lim}} \Phi(m,z)\,dm dz
\end{equation}
to the relevant luminosity dictated by the limiting magnitude of the data at each redshift.
In calculating each value of $M_{\rm lim}$,  
we apply a median K-correction for galaxies in each filter at each redshift calculated from the 
KCORRECT code \citep{BlantonRoweis2007} and using the galaxy templates of \cite{Kinney1996}.

In our data sample, we adjust the galaxy magnitudes for 
extinction using galaxy extinction E(B-V) values acquired from the 
NASA/IPAC Infrared Science Archive for each quasar, and assuming $R_{V}=3.1$ and the reddening curve of
\cite{ODonnell1994}.
In our calculations, we used a radius of 0.5 \hinv\ Mpc around the sample quasars and for the control fields derived from the
other 90Prime images of comparable depth in the same filter, and we fix the parameter $\gamma$ to be 1.77.
We follow \cite{Finn2001} and \cite{McLure2001} in using only galaxies in the range m*-1 - m*+2 
in order to eliminate many galaxies in the foreground or background of the quasar of interest.

\section{Results}
\subsection{Comparisons with other work}
\subsubsection{Correlation Amplitude}
\label{sec-bgqcomp}
\begin{figure}
\epsscale{1.0}
\plotone{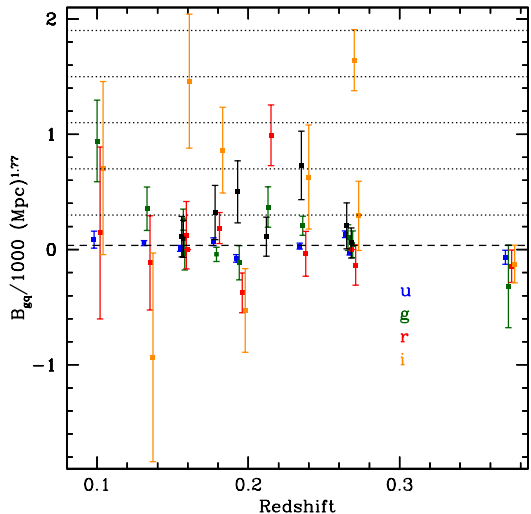}
\caption{
\bgq versus quasar redshift for all filters.  Points have been offset slightly in z for clarity.
Black points show values from the literature
listed in Table~\ref{tab-bgq}. Dotted lines, bottom to top, mark the \bgq\ values for Abell classes 0-4 \citep{YeeLopezCruz1999}
and the dashed line marks the galaxy-galaxy correlation value from \cite{DavisPeebles1983}.
\label{fig-bgqzall}}
\end{figure}

Results of the calculations of \bgq\ are listed in Table~\ref{tab-bgq} and plotted versus quasar redshift in Figure~\ref{fig-bgqzall},
along with values for some of our fields present in the literature.  We do not include a value for PG0844+349, as its low redshift, 
$z=0.064$, results in an angular radius corresponding to 0.5 \hinv\ Mpc that is larger than our contiguous chip areas.  
We do include this object in our discussion of the angular correlation function below.
We also omit the $i$ band data of Q2141+175 from both this calculation and the correlation function calculations as the 
quasar was placed in amp 4, which was masked due to noise. 

The limiting magnitudes of the galaxies included in these calculations are those listed in Table~\ref{tab-mlim} with the exception of
PG0953+414 $u$ and $i$ bands (23.7 and 23.2, respectively), PG1404+226 $i$ (22.5), and PG1612+261 $g$ (23.5), as there were no galaxies 
found in the control fields in these bands down to the quoted magnitude limits. 
This change in the magnitude limits had no effect on $N_t$, the quasar field galaxy counts, but simply increased $N_b$, the  
median number of background galaxies found in the control fields to a number greater than zero, resulting in a finite value of \bgq.
No significant differences in the \bgq\ values were found when imposing a uniform magnitude limit on all the fields, so we 
opt for using the individual values for each field.
All \bgq\ values
are quoted in units of Mpc$^{\gamma}$, where $\gamma=1.77$ unless otherwise noted.
The mean values of \bgq\ we find in ($u$, $g$, $r$, $i$) are (21$\pm$20, 161$\pm$96, 58$\pm$100, and 445$\pm$271).

Given the well-known morphology-density relation \citep{Dressler1980}
and the fact that radio loud quasars are more commonly found in massive early-type hosts \citep{Hamilton2002,Best2005},
it may be expected that galaxies cluster more strongly around radio loud quasars than around radio quiet objects.
Optical and X-ray selected AGN show no tendency to reside in galaxy group environments
\citep{Coldwell2003,Silverman2009} but \cite{Wold2001} find that radio loud quasars do prefer group or Abell class 0 environments.
Some other studies have found this distinction between the environments of
radio loud and radio quiet quasars  \citep{Ellingson1991,Best2005,Shen2009,Hickox2009,RamosAlmeida2013}
while others  have not \citep{Fisher1996,McLure2001,Finn2001,Wold2001,Coldwell2002}.
As we have only 2 radio loud quasars in our sample, we have little leverage to address this question. However,
excluding the two radio loud objects in our sample, PG1545+210 and Q2141+175, we find a systematic decrease 
in all four filters: $<$ \bgq $>=$ 
(9$\pm$18, 144$\pm$114, -39$\pm$56, 295$\pm$260).  
Removing the two additional $u$ fields (PG1444+407 and HS0624+690) and one $g$ field (HS0624+690) that do not 
have limiting magnitudes that reach M$^{*}$+2 at the redshifts of the AGN does not affect these means significantly, 
bringing them to 32$\pm$14 and 202$\pm$113 respectively.
The error bars reflect the standard deviation in the measurements among all fields in each filter.
The median redshift of our sample AGN is 0.18 and the median M$_{r}$ is -23.7.
These modified \bgq\ averages are all consistent
with the galaxy-galaxy correlation amplitude, 20 (\hinv  Mpc)$^{1.77}$ \citep{DavisPeebles1983},
although within the uncertainties, the average values in $g$ and $i$ are also consistent with richer environments,
as discussed in more detail for individual fields below. 

\cite{Finn2001} used images of quasar fields taken with the
F606W, F675W, and F702W filters and the 2.5\arcmin\ field of view of the Wide Field Planetary Camera 2 on the
{\it Hubble Space Telescope} and, scaling to our assumed value of H$_{0}$, find $<$\bgq $>=104\pm22$, $-33\pm27$, and $133\pm44$ in those respective
filters
for radio quiet quasars at median redshifts of 0.17, 0.42 and 0.4.
\cite{McLure2001} find $<$\bgq $>=326\pm94$ for their sample of radio quiet objects with median redshift of 0.16, 
also derived from  F606W and F675W {\it HST}/WFPC2 images.  Their dataset is a superset of that of \cite{Fisher1996} who found
$<$\bgq $>=132\pm37$.
These {\it HST} studies were limited to radii of $\sim$200 kpc, while ground-based studies typically extend to the same 0.5 \hinv\ Mpc scale as ours does.
The ground-based V, R, and I photometry of \cite{Wold2001} 
targets AGN at higher redshift than our AGN sample, $z_{\rm med}=0.714$, and they find  $<$\bgq $>=210\pm82$ for radio
quiet quasars.
All of these studies show average \bgq\ values that are larger than \bgg\ at the $2-3\sigma$ level.  Our overall results are more in line with
the earlier studies of 
\cite{Smith1995} and \cite{Ellingson1991}, 
both with $z_{\rm med}\sim0.4$ who found average \bgq\ values consistent with \bgg\ for radio quiet objects.

\begin{deluxetable}{lccc}
\tablecaption{B$_{gq}$ Measurements
\label{tab-bgq}}
\tablewidth{0pt}
\tablehead{
\colhead{QSO} &\colhead{B$_{gq}$ (Mpc$^{\gamma}$) \tablenotemark{1}}  &\colhead{Filter} &\colhead{Ref.\tablenotemark{2}}
}
\startdata
MRK586	         &-3$\pm$6           &$u$  &1 \\
                 &-36$\pm$137        &$g$  &1 \\
                 &0\ \tablenotemark{3} &$r$  &1 \\
                 &93$\pm$161 &F606W  &2 \\
                 &13$\pm$82  &F606W  &3 \\
HS0624+6907	 &-66$\pm$52         &$u$  &1 \\
                 &-319$\pm$279       &$g$  &1 \\
                 &-144$\pm$127       &$r$  &1 \\
                 &-126$\pm$151       &$i$  &1 \\
PG0923+201	 &-76$\pm$25           &$u$  &1 \\
                 &-114$\pm$133         &$g$  &1 \\
                 &-377$\pm$130         &$r$  &1 \\
                 &-529$\pm$291         &$i$  &1 \\ 
                 &500$\pm$269 &F675W  &2\\
                 &119$\pm$81  &F606W  &3\\
PG0953+414	 &30$\pm$28           &$u$ &1 \\
                 &207$\pm$101          &$g$ &1 \\
                 &-36$\pm$191         &$r$ &1 \\
                 &629$\pm$519         &$i$ &1 \\ 
                 &730$\pm$297 &F675W  &2\\
                 &159$\pm$81  &F606W  &3\\
PG1116+215	 &68$\pm$41            &$u$  &1 \\
                 &-41$\pm$57           &$g$  &1 \\
                 &186$\pm$134          &$r$  &1 \\
                 &863$\pm$461          &$i$  &1 \\ 
                 &321$\pm$235 &F606W   &2\\
                 &185$\pm$81  &F606W   &3\\
PG1307+085	 &10$\pm$27           &$u$  &1 \\
                 &259$\pm$120          &$g$  &1 \\
                 &125$\pm$306         &$r$  &1 \\
                 &1461$\pm$778        &$i$  &1 \\
                 &112$\pm$176 &F606W  &2\\
                 &170$\pm$82  &F606W  &3\\
PG1404+226	 &85$\pm$85           &$u$  &1 \\
                 &942$\pm$522         &$g$  &1 \\
                 &143$\pm$759         &$r$  &1 \\
                 &707$\pm$854         &$i$  &1 \\ 
PG1444+407	 &-27$\pm$19          &$u$  &1 \\
                 &46$\pm$119          &$g$  &1 \\
                 &-135$\pm$161        &$r$  &1 \\
                 &292$\pm$334         &$i$  &1 \\ 
                 &57$\pm$130 &F606W   &2\\
                 &49$\pm$79  &F606W   &3\\
PG1545+210	 &133$\pm$38          &$u$  &1 \\
                 &104$\pm$120         &$g$  &1 \\
                 &0\ \tablenotemark{3}  &$r$  &1 \\
                 &1643$\pm$404        &$i$  &1 \\
                 &206$\pm$199 &F606W  &2\\
                 &242$\pm$79  &F606W  &3\\
                 &236$\pm$120         &$r$ &4\\
PG1612+261	 &56$\pm$27           &$u$ &1 \\
                 &354$\pm$237         &$g$ &1 \\
                 &-116$\pm$397        &$r$ &1 \\
                 &-935$\pm$727        &$i$ &1 \\ 
Q2141+175        &370$\pm$221         &$g$ &1 \\
                 &991$\pm$355         &$r$ &1 \\
                 &112$\pm$169 &F675W  &2
\enddata
\tablenotetext{1}{Here we quote results for $\gamma=1.77$ and a radius of 0.5 \hinv Mpc; 
Literature values have been
scaled by $(h^{-1})^{1.77}$, but are otherwise uncorrected for different cosmologies.}
\tablenotetext{2}{
References: (1)this work; (2)\cite{McLure2001}; (3)\cite{Finn2001}; (4)\cite{Yee1984} }
\tablenotetext{3}{$N_{\rm t}=N_{\rm b}$}
\end{deluxetable}

Considering the individual values, we find that there can be a large range of \bgq\ values obtained for the same field depending on the 
filter used.
Typically, the values of \bgq\ in the $i$ band tend to be larger than those of the other bands and those calculated in the $u$ band are 
nearly all consistent with \bgg.  As above, for the $r$ band, we compare our individual results with those found in the literature, from 
\cite{Finn2001} and \cite{McLure2001},
and for PG1545+210, we compare with \cite{Yee1984}, who calculated \bgq\ from
$r$ band images taken with the
SIT-vidicon camera and the 1.52 m Palomar telescope.
We scaled these literature \bgq\ values for our value of H$_{0}$ but otherwise
do not correct for different cosmologies used in these other works
(\cite{McLure2001}: $H_0=50$ km s$^{-1}$ Mpc$^{-1}$; \cite{Finn2001} $H_0=100$ km s$^{-1}$ Mpc$^{-1}$, both
use $\Omega_{\rm m}=1$, $\Omega_{\Lambda}=0$).
In most cases our results are consistent with these previous studies.  We give a more detailed discussion of these 
comparisons for each field in Appendix~\ref{sec-notes}.

\cite{Yee1993} considered
\bgq $> 500$ to be a rich cluster environment, and clusters with 
Abell classes 0, 1 and 2 show values of $\sim$350, $\sim$650, and $\sim$950 respectively, according to \cite{Yates1989}.
\cite{McLure2001} suggest a classification based on that of \cite{YeeLopezCruz1999}, where
\bgq =(300,700,1100,1500,1900,2300) corresponds to Abell class (0,1,2,3,4,5) respectively.
Given the large statistical uncertainties in the \bgq\ values, we find results 
marginally greater than 500 in only a few cases, for PG1307+085 and PG1545+210 in the $i$ band, 1461$\pm$778 and 
1643$\pm$404,
respectively, 
for PG1404+226 in the $g$ band, 942$\pm$522, and
for Q2141+175 in the r band, 991$\pm$355. 
The radio loud AGN PG1545+210 is known to reside in an 
Abell class 1 environment \citep{Oemler1972}, though our $i$ band \bgq\ value is consistent with Abell class 3. The \bgq\ values
for this field in $g$ and $r$ are consistent with \bgg, and in $u$ we find only 133$\pm$38, 
which is significantly larger than \bgg, but 
far smaller than the $i$ band value.  
PG1307+085 has a foreground galaxy cluster in its field at a close projected separation from the 
quasar and with a similar redshift, $z_{\rm clus}=0.14$ versus $z_{q}=0.155$; and
PG1404+226 
has a background cluster within the 13.1\arcmin\ radius corresponding to 0.5 \hinv\ Mpc.
Q2141+175 is the only other radio loud AGN in our sample and while it
has no known cluster at its redshift or in projection, it is a less well-studied field than many of the others.
Thus the \bgq\ for these other fields may be influenced by foreground, background, or even proximate 
galaxy clusters, but only in select bands.

There are a handful of other fields that show a significant, positive signal in \bgq\ in one or more bands,
and these are discussed in Appendix~\ref{sec-notes}.  One field, PG0923+201, shows negative values of \bgq\
in all bands, with two ($u,r$) at $>2\sigma$. As we discuss in Appendix~\ref{sec-notes}, this well-studied field
has a compact group consistent with the redshift of the AGN and with a center only 23 \hinv\ kpc from the
position of the AGN, making this result an intriguing one.  Three other fields, HS0624+690, MRK586, and PG1444+407
show \bgq\ values in all filters that are either consistent with or significantly less than \bgg.  HS0624+690 is the
only quasar in our sample, and with $z=0.370$, our limiting magnitudes in $u$ and $g$ allow us to
reach only  $\sim M^{*}+0.7$ and $\sim M^{*}+0.1$, respectively.

Two other AGN in our sample, PG1116+215 and PG1612+261 
lie near galaxy clusters, though their redshifts and position on the sky place them near the peripheries of
those clusters. The \bgq\ values for these fields give no indication of a cluster association:
for PG1116+215, all the \bgq\ values are consistent with \bgg, except the anomalously high value we find for the 
$i$ band, 863$\pm$461.
For PG1612+261, the only filter showing a \bgq\ marginally larger than \bgg\ is the $g$ band, for which we find
354$\pm$237, Abell class 0.

\subsubsection{Correlation Functions}
\begin{figure}
\epsscale{1.0}
\plotone{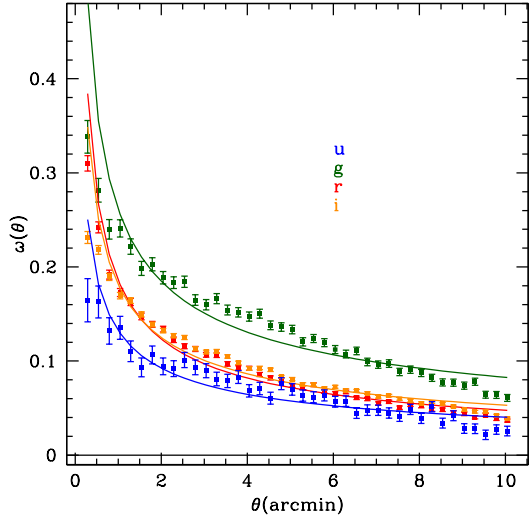}
\caption{
Galaxy-galaxy angular correlation function in $u,g,r,i$ filters for 90Prime fields.
\label{fig-gg}}
\end{figure}

We show the results for the normalization, $A_{\rm gg }$, and slope, $\gamma$ of the 
galaxy autocorrelation function in Figure~\ref{fig-gg} and list the results in Table~\ref{tab-gg}.
These calculations were performed for Sample 2.
For the $u$ and $g$ filters, we use only chips 1 and 4 as chips 2 and 3 added extra structure due to masking.
The $r$ and $i$ filters typically are less affected by this.  
The use of all chips 
for $u$ and $g$ gives similar results, but a poorer fit to the form of the autocorrelation function.
We avoid separations less than 10\arcsec\ to mitigate against blending, and because we are 
restricted to single chips, we calculate the correlation function to a maximum separation of ~10\arcmin.
The errors on the parameters are derived from the variance in that parameter from 100 jackknife samples of the dataset, where
the sampling with replacement was done one field/filter at a time.
\begin{deluxetable}{llll}
\tablecaption{Galaxy-Galaxy Clustering Clustering in 90Prime QSO fields\tablenotemark{1}
\label{tab-gg}}
\tablewidth{0pt}
\tablehead{
\colhead{Filter} &\colhead{$\gamma$}  &\colhead{$A_{\rm gg }$}  &\colhead{$\chi_{\nu}$} 
}
\startdata
$u$\ \tablenotemark{2}      &1.51$\pm$0.02  &0.13$\pm$0.05  &2.9   \\
$g$\ \tablenotemark{2}      &1.50$\pm$0.03  &0.26$\pm$0.08  &10.9 \\
$r$      &1.59$\pm$0.03  &0.18$\pm$0.04   &12.5  \\
$i$      &1.53$\pm$0.02  &0.18$\pm$0.03   &27.1  \\
\enddata
\tablenotetext{1}{0.1667\arcmin - 10.1667\arcmin}
\tablenotetext{2}{Excludes chips with one or more amplifiers masked}
\end{deluxetable}

\begin{figure}
\epsscale{1.0}
\plotone{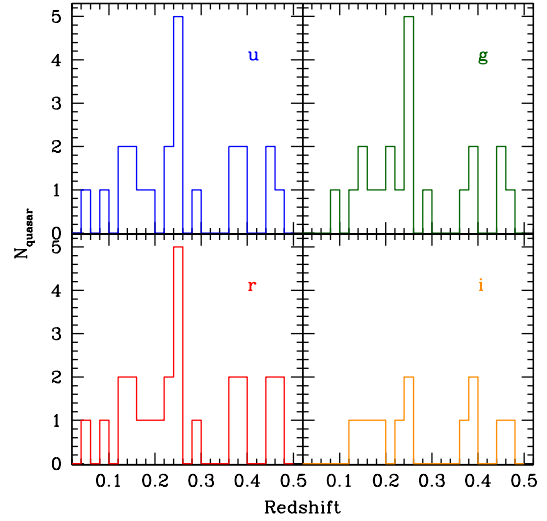}
\caption{Redshift distributions of all quasars in the 90Prime fields
\label{fig-zqso}}
\end{figure}

In our calculations of the galaxy-quasar cross-correlations, we consider quasars in 90Prime fields other than
the target quasars themselves, as any clustering may also be present for these objects as well.
As discussed above, the galaxy counts in our images are very sparse at $z>1$ and so galaxies associated with clusters, groups, or
quasars at these redshifts are generally not of concern to us here.
AGN with $z<0.5$ found in the 90Prime frames are listed in Table~\ref{tab-allq}
and the resulting quasar redshift distributions in each filter are shown in Figure~\ref{fig-zqso}. 
The absolute magnitudes relative to M$^{*}$ reached for $z=0.5$ galaxies in Samples 1 and 2 are listed in Table~\ref{tab-sample}.
\begin{deluxetable}{lcl}
\tablecaption{Other $z<0.5$ Quasars in 90Prime Fields
\label{tab-allq}}
\tablewidth{0pt}
\tablehead{
\colhead{QSO} &\colhead{z}  &\colhead{Filters} 
}
\startdata
\multicolumn{3}{c}{PG0844+349}  \\
\hline
SDSS J084731.77+351416.4 &0.237 &$u,r$ \\
2MASX J08503620+3455231  &0.144 &$u,r$ \\
\hline
\multicolumn{3}{c}{PG0923+201} \\
\hline
SDSS J092507.72+203540.9 &0.472  &$g,i,r$ \\
SDSS J092525.16+202139.0 &0.460  &$u,g,r$ \\
SDSS J092536.08+201649.5 &0.228  &$g$ \\
\hline
\multicolumn{3}{c}{PG1307+085} \\
\hline
Q87GBBWE91 1307+0843     &\nodata  &$r,i$ \\
SDSS J131155.76+085340.9 &0.469    &$u,g,r,i$ \\
\hline
\multicolumn{3}{c}{PG1444+407} \\
\hline
SDSS J144618.08+412003.0 &0.268  &$u,g,r,i$ \\
\hline
\multicolumn{3}{c}{PG1545+210} \\
\hline
SDSS J154749.70+205056.5 &0.265  &$u,g,r$ \\
SDSS J154750.71+210351.1 &0.296  &$u,g,r$ \\
SDSS J155014.81+212431.5 &0.479  &$r$ \\
SDSS J155046.30+205803.1 &0.401  &$u,g,r,i$ \\
2MASS J15505930+2128088  &0.372  &$u,r$ \\
\hline
\multicolumn{3}{c}{PG1612+261} \\
\hline
HB89 1612+266            &0.395   &$u,g,r,i$ \\
\enddata
\end{deluxetable}

\begin{figure}
\epsscale{1.0}
\plottwo{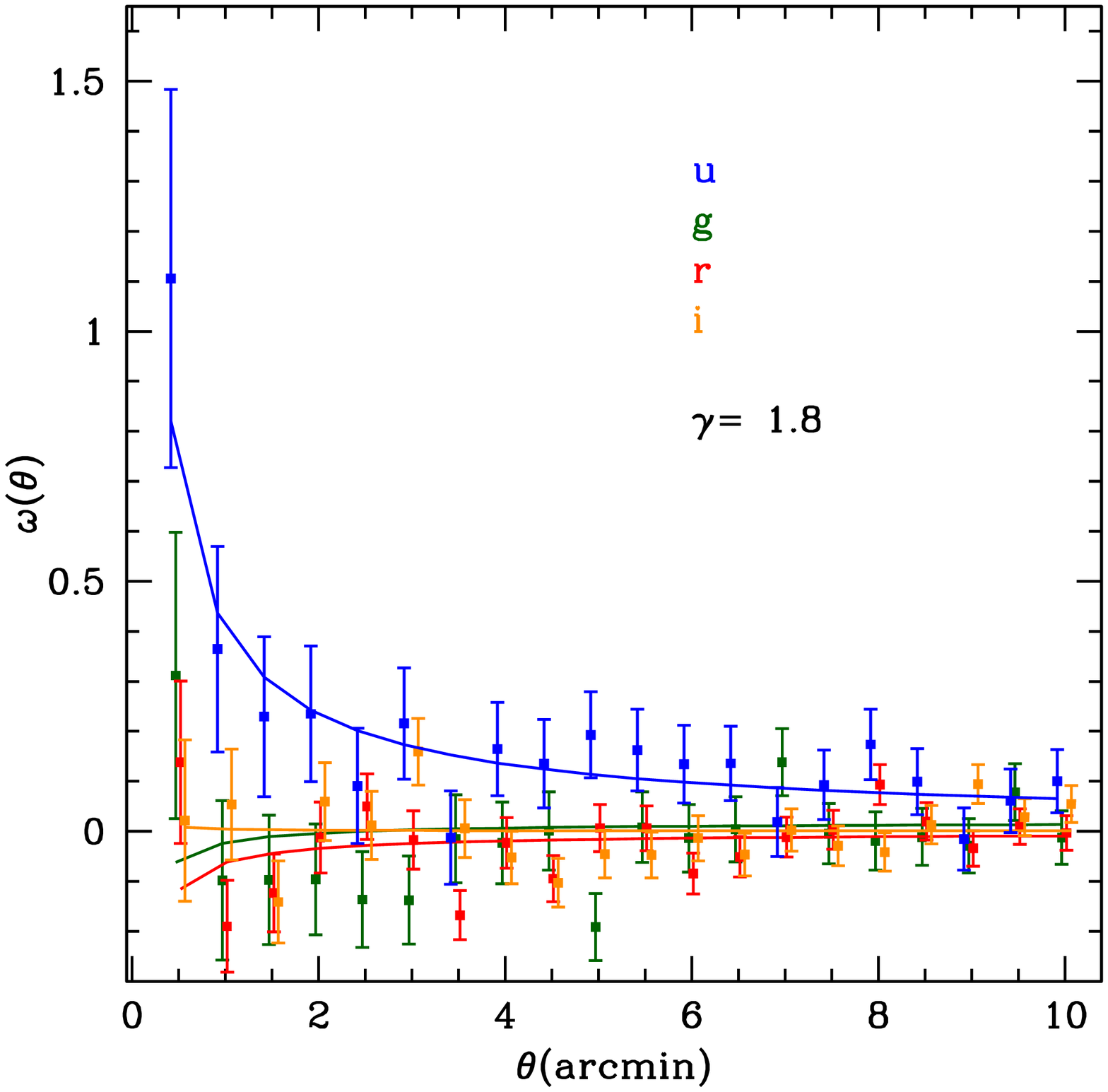}{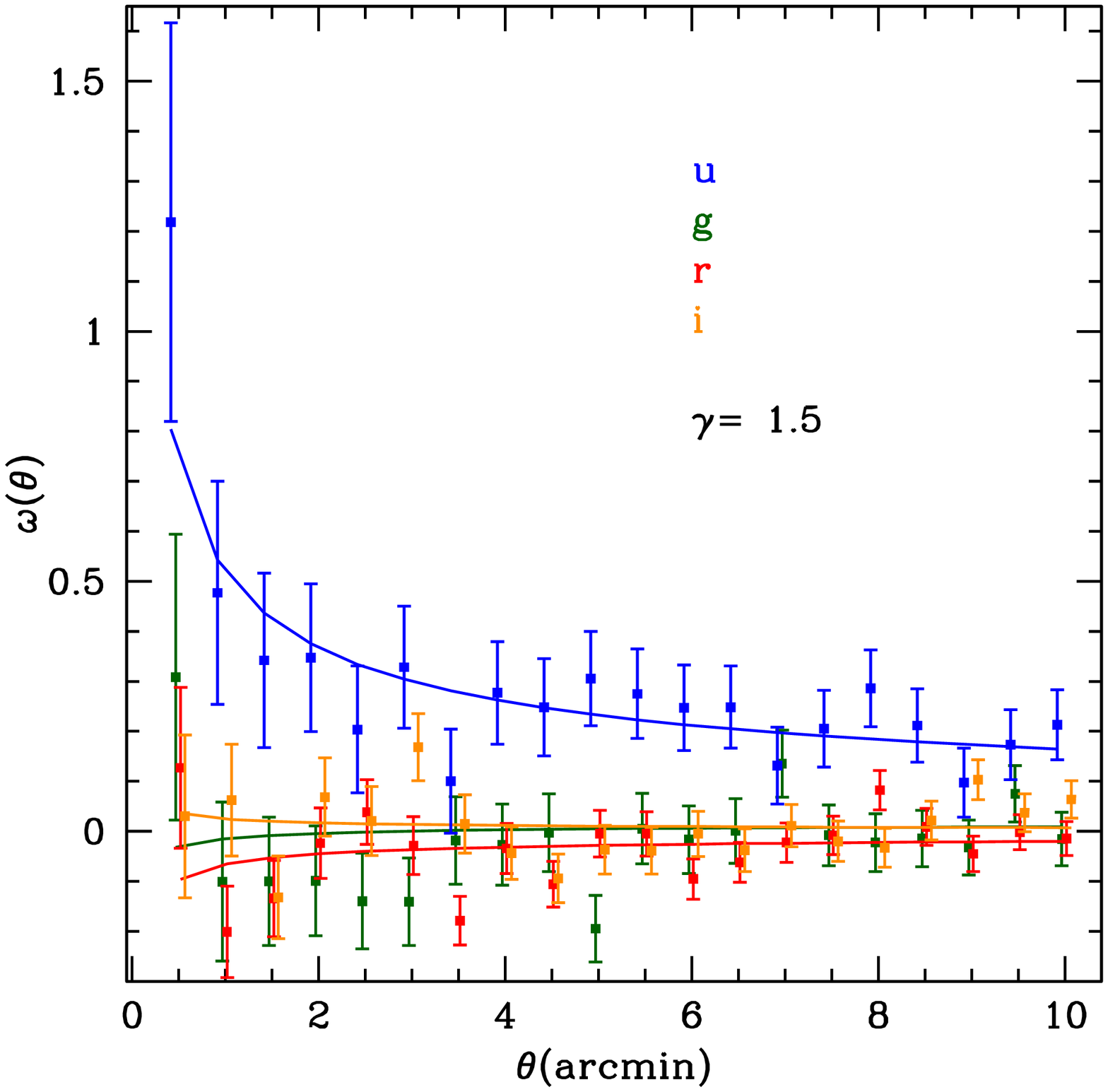}
\caption{
Quasar-galaxy angular correlation functions in $u,g,r,i$ filters for 90Prime fields for
$\gamma=1.8$ and $\gamma=1.5$.
Points are offset in x and g fit curve is offset in y by +0.02 for clarity.
\label{fig-qg}}
\end{figure}

We list the results for the normalization \agq\ of the galaxy-quasar cross-correlation function in Table~\ref{tab-gq}
for both Samples 1 and 2, where the latter correspond to deeper magnitude limits and fainter galaxy samples.
For these solutions, we hold the slope $\gamma$ to be constant at 1.8, the commonly accepted value and 1.5, corresponding
to the slope we found from the 90Prime data for the galaxy-galaxy clustering on 10\arcmin\ scales.  This difference in slope
changes the normalization at only the $\sim1\sigma$ level.
The results for
both slopes are plotted in Figure~\ref{fig-qg}.	
\begin{deluxetable}{lcll}
\tablecaption{Galaxy-Quasar Clustering in 90Prime QSO fields\tablenotemark{1}
\label{tab-gq}}
\tablewidth{0pt}
\tablehead{
\colhead{Filter} &\colhead{$A_{\rm gq }$}  &\colhead{$\chi_{\nu}$} &\colhead{r$_0$ h$^{-1}$ Mpc\tablenotemark{2}}
}
\startdata
\multicolumn{4}{c}{Sample 1, $\gamma=1.8$} \\
\hline
$u$      &0.49$^{+0.15}_{-0.14}$  &0.39   &8.07 \\
$g$      &0.23$^{+0.16}_{-0.14}$  &0.69   &5.36 \\
$r$      &0.21$^{+0.08}_{-0.07}$  &0.70   &6.11\\
$i$      &-0.03$\pm0.07$          &1.66   &\nodata \\
\hline
\multicolumn{4}{c}{Sample 1, $\gamma=1.5$} \\
\hline
$u$      &0.61$^{+0.21}_{-0.19}$    &0.38  &16.9\\
$g$      &0.34$^{+0.22}_{-0.19}$    &0.65  &11.64\\
$r$      &0.28$\pm$0.10             &0.65  &12.3\\
$i$      &-0.0027$^{+0.10}_{-0.09}$ &1.68  &\nodata\\
\hline
\multicolumn{4}{c}{Sample 2, $\gamma=1.8$} \\
\hline
$u$      &0.40$^{+0.12}_{-0.11}$   &0.61   &7.76\\ 
$g$      &-0.04$\pm$0.09  &1.03   &\nodata\\ 
$r$      &-0.05$\pm$0.05  &1.57   &\nodata\\
$i$      &0.004$\pm$0.05  &1.44   &0.90\\
\hline
\multicolumn{4}{c}{Sample 2, $\gamma=1.5$} \\
\hline
$u$      &0.51$^{+0.16}_{-0.15}$   &0.57 &16.2\\
$g$      &-0.03$^{+0.12}_{-0.11}$  &1.04 &\nodata \\
$r$      &-0.06$^{+0.07}_{-0.06}$  &1.58 &\nodata\\ 
$i$      &0.02$\pm$0.07            &1.44 &2.95\\ 
\enddata
\tablenotetext{1}{0.1667\arcmin - 10.1667\arcmin}
\tablenotetext{2}{$\epsilon=0$}
\end{deluxetable}

\cite{CroomShanks1999} found no galaxy-quasar correlation in their sample of $\sim$150 optically and X-ray selected quasars
with $z=1-1.5$ and $b_{J}<23$ galaxies in five deep AAT plate fields.  In fact, their data showed a weak anti-correlation between these
two populations.
At the the lower redshifts of our sample, we see significant correlations only in the 
$u$ band.  For Sample 1, there is also a significant positive correlation
on the $r$ band, but this does not hold with the inclusion of fainter galaxies in Sample 2.  In fact in all bands except the $i$ band, 
the correlation
is significantly smaller for Sample 2, and we also see weak anti-correlations in several cases. In $i$, the normalizations
are all consistent with zero for both Samples 1 and 2.

In Table~\ref{tab-gq}, we also tabulate the correlation length for galaxy-quasar clustering, 
$r_{0}$, found from Limber's equation, Equation~\ref{equ-limber}, and the galaxy and quasar
redshift distributions plotted in Figures~\ref{fig-zdist} and \ref{fig-zqso}. This is undefined 
for negative values of \agq, so no correlation length values are reported in Table~\ref{tab-gq} in these cases.
The values we find range from 0.9 \hinv\ Mpc  for Sample 2 in the $i$ band with $\gamma=1.8$ to 
$\sim$16 \hinv\ Mpc for both Samples 1 and 2 in the $u$ band with $\gamma=1.5$. These results are generally within the range 
found for $0.7<z<1.4$ quasars in the DEEP2 survey by
\cite{Coil2007}, 0.1 \hinv\ $\lesssim r_0  \lesssim$ 10 \hinv\ Mpc.
\cite{Brown2001} studied a population of galaxy fields around a sample of moderate redshift quasars with $z=0.2-0.7$ and also
find a weak correlation, though a stronger one for red galaxies ($r_0 \sim 7$ \hinv\ Mpc, $\gamma \sim 1.9$) than for blue galaxies
($r_0 \lesssim 4$ \hinv\ Mpc).
These authors provide a summary of results to that time in their Table 1.
The value of the correlation length ranges from 
$\sim$5-8  \hinv\ Mpc for samples of radio quiet quasars with similar redshift but shallower depth than our current 90Prime study
to $\sim$12-17 \hinv\ Mpc for fields around radio loud quasars, comparable to our result for the $u$ band in the 90Prime sample.
Below where we discuss trends of galaxy-quasar clustering  with
quasar properties, we look at the question of whether the two radio loud objects in our sample show significantly
greater galaxy clustering than the radio quiet objects.

\begin{figure}
\epsscale{1.0}
\plotone{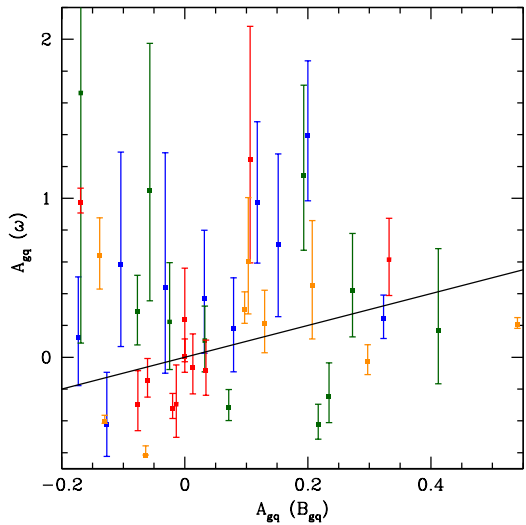}
\caption{
Comparison of the two methods for calculating the angular covariance amplitude, \agq\, from
Equation~\ref{equ-omegagq} and Equation~\ref{equ-agq}, both with $\gamma=1.8$. 
\label{fig-agq}}
\end{figure}

In our two methods of investigating galaxy-quasar correlations, outlined in Sections~\ref{sec-omegagq} and \ref{sec-bgq},
we calculate the angular covariance amplitude, \agq\,
in two different ways, one involving a comparison with a set of random galaxy positions and one using the
expectation from an integral of the galaxy luminosity function in the relevant band.  In Figure~\ref{fig-agq}, we
compare the values obtained from these two methods for $\gamma=1.8$.  We note that the uncertainties in the fitted
$\omega (\theta)$ for a single field/filter can be large, and 
the offset for any one field/filter can be large, ranging from
-0.6 to +1.8, but the mean offset in these values is 0.22
with a standard deviation of 0.55.
As has been noted in  previous work, clustering statistics like \bgq\  can be sensitive to the methodology used to calculate it,
the choice of luminosity function and control fields.  We therefore concentrate less on the absolute values
of this parameter than on investigating any trends in the data, discussed in the following section.

\subsection{Trends in Clustering Parameters}
\subsubsection{Correlation Amplitude}
We investigate trends of galaxy-quasar clustering with quasar absolute $r$ magnitude, black hole mass, and [\ion{O}{3}] line luminosity,
as an indicator of AGN activity \citep{Kauffmann2004}.
These properties are listed in Table~\ref{tab-quasars}.  
The values of M$_{r}$ are derived from the $r$ listed in the SDSS DR9 catalog, with Galactic extinction calculated using the
reddening values from
\cite{Schlafly2011} and the 
\cite{ODonnell1994} reddening law. We also applied a K-correction, calculated from the QSO composite spectrum of 
\cite{VandenBerk2001} and the SDSS filter response curves.
We collected black hole mass and [\ion{O}{3}] $\lambda$5007 line luminosity measurements from the literature 
and list them in Table~\ref{tab-quasars}.
Where the [\ion{O}{3}] line luminosities do not exist in the literature, we
downloaded the SDSS DR10 spectra for these objects and fit the H$\beta$ [\ion{O}{3}]$\lambda$4959,5007 complex using the
IRAF task {\tt specfit} and list the resulting value in Table~\ref{tab-quasars}.  
\begin{deluxetable}{llccll}
\tablecaption{Sample Quasar Properties
\label{tab-quasars}}
\tablewidth{0pt}
\tablehead{
\colhead{QSO} &\colhead{M$_{r}$} &\colhead{log(M$_{\rm BH}$)} &\colhead{Refn.\tablenotemark{1}}&
\colhead{[\ion{O}{3}] Luminosity} & \colhead{Refn. \tablenotemark{2}}\\
\colhead{} & \colhead{} &\colhead{}  &\colhead{} &\colhead{erg/s $\times10^{42}$} & \colhead{}
}
\startdata
MRK586	     &-23.4 &8.34 &1 &3.50 &1 \\
HS0624+690   &-26.0 &9.69 &2 &... &... \\
PG0844+349   &-22.4 &7.96 &3 &... &...\\
PG0923+201   &-23.7 &8.00 &4 &0.99 &2 \\
PG0953+414   &-24.7 &8.44 &3 &1.90 &3 \\
PG1116+215   &-24.6 &8.52 &4 &2.60 &3 \\
PG1307+085   &-23.2 &8.64 &3 &3.17 &2 \\
PG1404+226   &-21.8 &6.88 &4 &0.38 &2 \\
PG1444+407   &-24.1 &8.28 &4 &2.22\tablenotemark{3} &2 \\
PG1545+210   &-24.4 &9.31 &4 &4.93 &2 \\
PG1612+261   &-22.6 &8.05 &4 &6.68\tablenotemark{4} &2 \\ 
Q2141+175    &-23.6 &8.74 &1 &2.80 &1 \\
\enddata
\tablenotetext{1}{References: (1)\cite{McLure2001b}; (2)\cite{Labita2006}; (3)\cite{Peterson2004}; (4)\cite{Vestergaard2006}}
\tablenotetext{2}{References: (1)\cite{HoKim2009}; (2) This study; (3)\cite{Shang2007}}
\tablenotetext{3}{\cite{Shang2007} find 0.22}
\tablenotetext{4}{Sum of two components FWHM=420 \kms and 1000 \kms with fluxes 4.75 and 1.93 respectively}
\end{deluxetable}

\begin{figure}
\epsscale{1.0}
\plotone{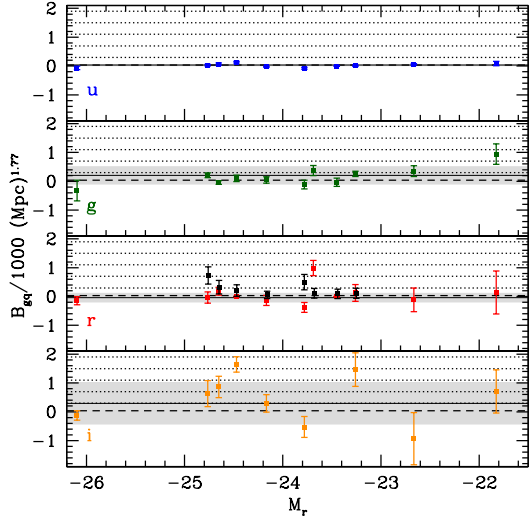}
\caption{
\bgq\ in each filter versus quasar absolute r magnitude.  Solid lines and shaded regions mark the mean and dispersion among all the
values in a given filter, dotted lines show the expected correlation amplitude for Abell classes 0-4 \citep{YeeLopezCruz1999}, and the
dashed line marks the galaxy-galaxy correlation value from \cite{DavisPeebles1983}.
Black points show values from the literature
listed in Table~\ref{tab-bgq}.
\label{fig-bgqm}}
\end{figure}
\begin{figure}
\epsscale{1.0}
\plotone{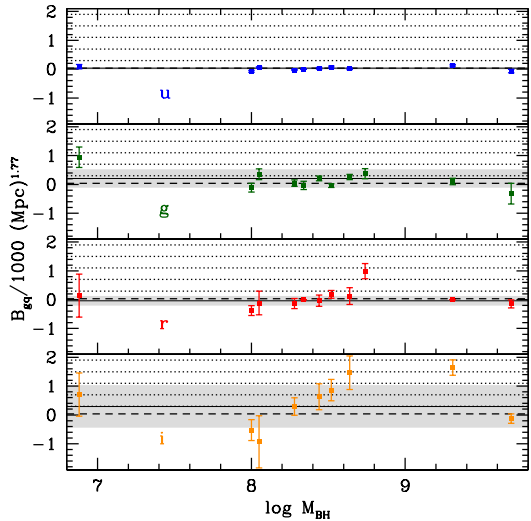}
\caption{As in Figure~\ref{fig-bgqm} but for
\bgq\ in each filter versus quasar black hole mass
\label{fig-bgqmbh}}
\end{figure}
\begin{figure}
\epsscale{1.0}
\plotone{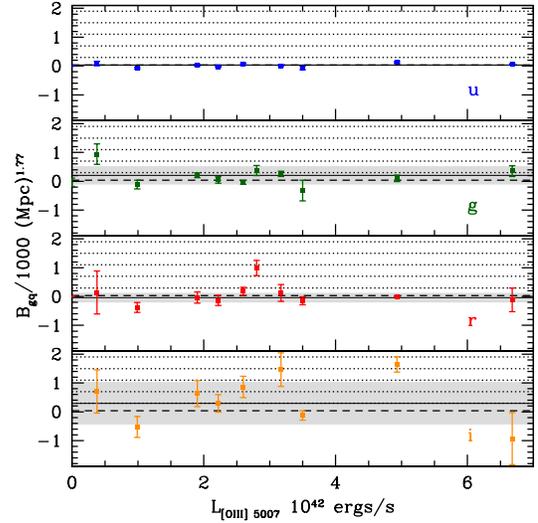}
\caption{As in Figure~\ref{fig-bgqm} but for
\bgq\ in each filter versus quasar [\ion{O}{3}] luminosity
\label{fig-bgqo3}}
\end{figure}

Plots of these quasar properties with \bgq\ parameter are shown in
Figures~\ref{fig-bgqm}-\ref{fig-bgqo3}, but in no case do we find a significant trend of this clustering measure with
any of the quasar properties.

\subsubsection{Correlation Functions}
We also investigate trends of the
galaxy-quasar angular correlation function with 
quasar luminosity, black hole mass, and [\ion{O}{3}] line luminosity by using
the relative bias with respect to galaxies, $b_{gq} = (\omega_{gq}/\omega_{gg})^{1/2}$.
Overall, we find $b_{gq}= (3.08\pm0.51, 1.49\pm0.53, 1.48\pm0.29, 0.95\pm1.49)$ for $(u, g, r, i)$ for cross- and 
autocorrelation function 
solutions with fixed $\gamma=1.5$. The bias is only significantly greater than unity for the $u$ band and decreases with increasing
wavelength, perhaps reflecting a 
trend seen for SDSS Sy 2 galaxies to lie in bluer environments than normal galaxies \citep{Coldwell2014}.
This runs counter to a trend in the mean \bgq\ values quoted above, for which the $u$ band values are lowest and those in $i$ are the largest.
This is driven by the large $i$ band \bgq\ values found  for PG1545+201 and PG1307+085, the former being a radio loud quasar.  
That trend is less apparent for the radio quiet objects alone, and without PG1307+085 as well, the average $i$ band \bgq\ value drops to
129$\pm$238.

\begin{deluxetable}{lccc}
\tablecaption{Galaxy-Galaxy and Galaxy-Quasar  Correlations for each 90Prime Field\tablenotemark{1}
\label{tab-omegarat}}
\tablewidth{0pt}
\tablehead{
\colhead{QSO} &\colhead{Filter} &\colhead{$\omega_{\rm gg}$} &\colhead{$\omega_{\rm gq}$}
}
\startdata
MRK586	      &$u$  &0.151$\pm$0.036 &0.08$\pm$1.08 \\
              &$g$  &0.015$\pm$0.008 &0.33$\pm$0.54 \\
              &$r$  &0.072$\pm$0.006 &0.07$\pm$0.47 \\
HS0624+6907   &$u$  &0.071$\pm$0.024 &-0.67$\pm$0.32 \\ 
              &$g$  &0.387$\pm$0.055 &0.67\tablenotemark{2}  \\
              &$r$  &0.041$\pm$0.008 &-0.50$\pm$0.24 \\
              &$i$  &0.082$\pm$0.008 &-0.98$\pm$0.01 \\
PG0844+349    &$u$  &0.215$\pm$0.022 &0.95$\pm$0.50 \\
              &$r$  &0.006$\pm$0.003 &-0.08$\pm$0.12 \\
PG0923+201    &$u$  &0.150$\pm$0.019 &0.19$\pm$0.53 \\
              &$g$  &0.194$\pm$0.010 &0.57$\pm$0.36 \\
              &$r$  &0.223$\pm$0.011 &0.95$\pm$0.32 \\
              &$i$  &0.168$\pm$0.005 &0.54$\pm$0.37 \\ 
PG0953+414    &$u$  &0.206$\pm$0.012 &0.32$\pm$0.47 \\
              &$g$  &0.344$\pm$0.020 &1.65$\pm$0.93 \\
              &$r$  &0.246$\pm$0.008 &-0.36$\pm$0.31 \\
              &$i$  &0.270$\pm$0.004 &0.19$\pm$0.30 \\ 
PG1116+215    &$u$  &0.433$\pm$0.028 &1.13$\pm$0.86 \\
              &$g$  &0.630$\pm$0.039 &1.80$\pm$1.40 \\
              &$r$  &0.252$\pm$0.017 &2.47$\pm$1.31 \\
              &$i$  &0.127$\pm$0.006 &0.83$\pm$0.61 \\ 
PG1307+085    &$u$  &0.094$\pm$0.023 &0.39$\pm$0.62 \\
              &$g$  &0.158$\pm$0.015 &0.42$\pm$0.53 \\
              &$r$  &0.088$\pm$0.007 &0.35$\pm$0.27 \\
              &$i$  &0.104$\pm$0.005 &0.01$\pm$0.19 \\
PG1404+226    &$u$  &0.126$\pm$0.013 &0.54$\pm$0.77 \\
              &$g$  &0.123$\pm$0.007 &0.24$\pm$0.87 \\
              &$r$  &0.171$\pm$0.004 &-0.15$\pm$0.28 \\
              &$i$  &0.170$\pm$0.005 &1.09$\pm$0.63 \\ 
PG1444+407    &$u$  &0.257$\pm$0.022 &0.36$\pm$0.96 \\
              &$g$  &0.029$\pm$0.008 &-0.03$\pm$0.30 \\
              &$r$  &0.159$\pm$0.007 &-0.35$\pm$0.20  \\
              &$i$  &0.115$\pm$0.005 &0.10$\pm$0.24  \\ 
PG1545+210    &$u$  &-0.029$\pm$0.014 &0.33$\pm$0.25 \\
              &$g$  &0.257$\pm$0.018  &-0.46$\pm$0.16 \\
              &$r$  &0.063$\pm$0.006  &-0.17$\pm$0.14 \\
              &$i$  &0.093$\pm$0.005  &0.09$\pm$0.21 \\
PG1612+261    &$u$  &0.178$\pm$0.023  &1.84$\pm$0.79 \\
              &$g$  &0.432$\pm$0.009  &-0.47$\pm$0.18  \\
              &$r$  &0.203$\pm$0.006  &-0.47$\pm$0.15 \\
              &$i$  &0.199$\pm$0.005  &-0.50$\pm$0.13 \\ 
Q2141+175     &$g$  &0.172$\pm$0.009  &-0.23$\pm$0.27  \\
              &$r$  &0.389$\pm$0.008  &0.17$\pm$0.39  \\
\enddata
\tablenotetext{1}{Calcuated within $0.1667\arcmin < \theta < 1.1667\arcmin$  for fixed $\gamma=1.6$}
\tablenotetext{2}{No galaxies in this inner bin, so formal error is indeterminate.}
\end{deluxetable}

The values of $\omega$ for each field and filter individually are listed in Table~\ref{tab-omegarat}.
For each field/filter where both the cross- and autocorrelation values are positive
in the innermost bin, $\omega<1.167\arcmin$,
we find no correlations in $b_{gq}$ with any of the quasar properties discussed above.
Most fields for which both $\omega_{gq}$ and $\omega_{gg}$ within 1.167\arcmin\ are positive show $b_{gq}>1$,  but only
a handful at $\gtrsim2\sigma$: PG0844+349 and PG1612+261 in $u$, HS0624+690 in $g$, PG0923+201 and PG1116+215 in $r$, and 
PG1404+226 in $i$.
For PG0844+349 and PG1612+261 the filters with $b_{gq}>1$ are the only filters with $\omega_{gq}>0$, for which
$b_{gq}$ is defined.  
That the $u$ filters would show a large positive relative bias while all others show a quasar-galaxy anticorrelation
may be indicative of bluer environments, although this does not hold for the HS0624+690 field, for which the $b_{gq}$ in  $g$ 
is $1.31\pm0.094$ but the values of $\omega_{gq}$ are negative in all other filters, including $u$.
A few fields show $b_{gq}>1$ consistently in all filters (PG0923+201, PG1116+215) or in all filters but one:
PG1307+085 ($i$), PG1404+226 ($r$).
The bias seen in the $u$ band for the radio loud quasar PG1545+210 is negative, and
not significantly
greater than unity in any filter.
Similarly, for the other radio loud AGN in our sample, Q2141+175, only $g$ and $r$ band data are available and neither
shows a large bias.  In $g$, $\omega_{gq}<0$, and in $r$ $b_{gq}=0.66\pm0.75$.

\section{Discussion}
A somewhat contradictory picture emerges when we compare the results of the correlation function analysis to the 
calculated \bgq\ statistics. 
The fields that show the largest values of \bgq,  PG1307+085, PG1545+210, PG0953+414 in the $i$ band, PG1404+226 in $g$, and Q2141+175 in $r$,
all show relative biases that are consistent with unity.
The PG0923+201 field showed an anticlustering \bgq\ signal in $u$ and $r$, but both filters give $b_{gq}>1$, with the $r$ band
significant at $\sim3\sigma$.
We note that, in addition to a luminosity function dependence, the \bgq\ analysis probes a larger spatial scale than the 
relative bias calculation within 1.167\arcmin, corresponding to $\sim$10-40\hinv\ kpc over the redshift range of the AGN in our sample.

Overall, as noted above in Section~\ref{sec-bgqcomp}, the mean values of the correlation amplitude for 0.5 \hinv\ Mpc scale clustering,
while all consistent within the uncertainties with galaxy-galaxy clustering, show the largest signal in the $i$ band, 
\bgq=295$\pm$260. The
correlation function analysis, on the other hand, probing $\sim$10-40\hinv\ kpc scales shows the largest bias for quasar galaxy clustering
in the $u$ band, with galaxy clustering around quasars a factor of 3 larger than the galaxy autocorrelation signal at these scales.   
This suggests that different galaxy types may cluster around quasars on different scales.  
Some earlier studies have seen clustering signals at small scales and differences in galaxy populations near quasars.
In a future paper, we will use the multi-band data to calculate photometric redshifts for the galaxies in our sample,
and investigate trends in the clustering parameters with galaxy types
determined with these redshifts.

The basic trends in our results are consistent with earlier findings that quasars cluster like L$^*$ galaxies, and indeed like galaxies 
up to 2 magnitudes fainter than L$^*$.
We also confirm that there are no significant
trends with quasar luminosity \citep{Finn2001,Serber2006,Coil2007}, though unlike \cite{Serber2006}, we also find that the
lack of luminosity dependence holds at small as well as large spatial scales.
In our sample of broad line AGN,
we do not find the trend for AGN with lower [\ion{O}{3}] luminosities
to have larger clustering signals as seen in SDSS narrow line AGN \citep{Wake2004}.

Thus, the mechanism driving accretion onto the central
black holes in these objects is not imprinted on their large scale galaxy environments.
This may be due to a non-merger mode in fueling the central black holes or to a delay in the onset of nuclear activity from the
time of the initial galaxy interactions which triggered its fueling.

\acknowledgements{ 
J.\ S.\  and A.\ R.\ acknowledge the support of the National Science Foundation (AST-0952923).

This research has made use of the NASA/IPAC Extragalactic Database (NED) which is operated by the Jet Propulsion Laboratory, California
Institute of Technology, under contract with the National Aeronautics and Space Administration.

Funding for the SDSS and SDSS-II has been provided by the Alfred P. Sloan Foundation, the Participating Institutions, the National Science 
Foundation, the U.S. Department of Energy, the National Aeronautics and Space Administration, the Japanese Monbukagakusho, the Max Planck 
Society, and the Higher Education Funding Council for England. The SDSS Web Site is http://www.sdss.org/.

The SDSS is managed by the Astrophysical Research Consortium for the Participating Institutions. The Participating Institutions are the 
American Museum of Natural History, Astrophysical Institute Potsdam, University of Basel, University of Cambridge, Case Western Reserve 
University, University of Chicago, Drexel University, Fermilab, the Institute for Advanced Study, the Japan Participation Group, Johns Hopkins 
University, the Joint Institute for Nuclear Astrophysics, the Kavli Institute for Particle Astrophysics and Cosmology, the Korean Scientist 
Group, the Chinese Academy of Sciences (LAMOST), Los Alamos National Laboratory, the Max-Planck-Institute for Astronomy (MPIA), the Max-
Planck-Institute for Astrophysics (MPA), New Mexico State University, Ohio State University, University of Pittsburgh, University of 
Portsmouth, Princeton University, the United States Naval Observatory, and the University of Washington.
}

Facilities: \facility{Bok}

\appendix
\section{Notes on Individual Objects}
\label{sec-notes}

In this section, we outline notes on each of our 90Prime fields, based on previous studies of the quasar fields and
host galaxies, searches for nearby objects in the NASA Extragalactic Database (NED), and the results of our
calculations of the correlation amplitude, \bgq.  We refer the reader to Table~\ref{tab-bgq} and Figures~\ref{fig-bgqzall}--\ref{fig-bgqo3}.

\subsection{Mrk 586, $z=0.155$
\label{sec-mrk586}}
The field of this radio quiet object has been imaged previously in narrow band [O~{\sc iii}] by 
\cite{Stockton1987} and in the optical (B,$g$,$i$ filters) by
\cite{Kirhakos1994} using the Palomar 1.5 m (image sizes $\sim$3\arcmin\ $\times$ 3\arcmin) who note 
four galaxies in this field, including a large disk galaxy (NAB 0205+02: 06)
at a projected separation of 15.6 \hinv\ kpc.

The quasar host was studied by \citep{Bahcall1997} using the {\it HST}/
Wide Field/Planetary Camera-2 (WFPC2) and the F606W filter.
The host is disk-like and the AGN has a
companion with jet-like structure at a projected distance of 23 \hinv\ kpc \citep{Stockton1987,Bahcall1997}.
There is no evidence of tidal features to suggest a direct interaction.

A NED search reveals $\sim$17 galaxy clusters, groups, or candidates, within 50\arcmin\ of the quasar. The cluster with the
smallest angular separation from the quasar (7.7\arcmin) is NSCS J020723+024645, which lies at $z=0.37$.

The \bgq\ we calculate from the 90Prime $r$ band image is identically zero because the number of galaxies detected at
5$\sigma$ within 0.5 \hinv\ Mpc of the quasar exactly equals the median number found in 871 control fields.  
This is 
consistent  with the values found by both \cite{McLure2001} and \cite{Finn2001} from the {\it HST}/WFPC2 images in the
F606W filter, 93$\pm$161 and 13$\pm$82, respectively, 
and with the values calculated from the 90Prime $u$ and $g$ band images, -3$\pm$6 and -36$\pm$137, respectively.
We have no 90Prime data of this field in the $i$ band.

\subsection{HS0624+690, $z=0.370$, QSO}
The host of this radio quiet quasar is an elliptical galaxy \citep{Floyd2004}.
\cite{Kirhakos1994} identify six galaxies in their ground-based optical images within the $\sim$3\arcmin\ $\times$ 3\arcmin\ frame.
From NED, we find that within 50\arcmin\ of the quasar, there is one known galaxy cluster, 
Abell 0557 (redshift unknown), at 8.03\arcmin\ separation from the quasar.

All the \bgq\ values we calculate for this field are less than zero,
indicating an underdensity of galaxies 
within 0.5 \hinv\ Mpc of the quasar, however, none of these underdensities is significantly less than \bgg\ at $>2\sigma$.
We conclude that the field around this object
is consistent with the field.
A caveat to this is that HS0624+690 is the the highest redshift AGN in our sample, and
the limiting magnitudes of the $u$ and $g$ coadded images are not significantly fainter than M$^{*}$ at the quasar redshift,
M$^{*}$ + 0.7 and M$^{*}$ + 0.1, respectively,
so we are not reaching as far down the luminosity function for this field as the others in the sample. 

\subsection{PG0844+349, $z=0.064$
\label{sec-pg0844}}
This radio quiet object has a host with bulge+disk structure, showing no bar, but some evidence of morphological disturbance
from deep H-band imaging
\citep{Guyon2006,Veilleux2009}.  
This object qualifies as having a bright interacting companion (H band luminosity greater than 1/10 L$_*$ and projected distance
within 24 \hinv\ kpc \cite{McLeod1994}, though it is not likely currently undergoing a strong interaction since
\cite{Veilleux2009} classify this object as a post-merger object with a single nucleus.
Also, from the [O~{\sc ii}] measurement of \cite{Wilkes1999}, \cite{Ho2005} estimates a star formation rate of the host galaxy of
less than 1.6 M$_{\odot}$ yr$^{-1}$.
Within a 50\arcmin\ radius around this quasar, NED returns 58 galaxy clusters, groups, or candidates.  

\subsection{PG0923+201, $z=0.192$}
The host of this radio quiet AGN is an elliptical galaxy with no evidence of a bar or of morphological disturbance from an interaction with 
a companion \citep{Dunlop2003,Guyon2006,Veilleux2009}.
Previous authors have note the presence of two bright galaxies within 16 \hinv\ kpc of the quasar and have 
suggested it is a member of a small group \citep{Heckman1984,McLeod1994,Bahcall1997}.
Indeed, the compact group SDSSCGB 30540
($z=0.19$) identified by \cite{McConnachie2009} in the SDSS DR6 
has a redshift consistent with that of PG0923+201. 
While the quasar is not listed as a group member in their catalog, it has a separation of $\sim$23 \hinv\ kpc from the group center.
There are $\sim$45 other galaxy clusters or groups within 50\arcmin\ of this quasar.

Our \bgq\ values from the $r$ and $i$ band frames, -377$\pm$130 and -529$\pm$291, are inconsistent with those of \cite{McLure2001}
and \cite{Finn2001} at the 3.5-4$\sigma$ level.  These authors find moderately significant overdensities in this field 
while we see underdensities in all filters, of increasing magnitude from $u$ to $i$. 
This underdensity is surprising if the quasar is in fact associated with SDSSCGB 30540.

\subsection{PG0953+414, $z=0.234$}
\cite{Dunlop2003} classify the host galaxy of this radio quiet AGN as an elliptical, while others \citep{Bahcall1997,Guyon2006} have found it
to show disk-like structure but to be too faint to classify reliably. \cite{Kirhakos1994} find 22 likely galaxies in their optical
images described above in Sec.~\ref{sec-mrk586}.

There are 54 galaxy groups, clusters or candidates within 50\arcmin\ of this quasar, one of which, GMBCG J149.20633+41.27869, has a projected
separation of only 1.455\arcmin, though it lies well behind PG0953+414, at $z=0.362$.
A second cluster, WHL J095714.9+411700, is separated by 4.532\arcmin\ from the quasar and has a redshift $z=0.446$.

Our $i$ band  result for \bgq\ is consistent with the result of \cite{McLure2001} using {\it HST}/WFPC2 F675W.
Both values are large, $\sim600-700$, but with large statistical uncertainty, $\sim300-400$.  
The $g$ band value for this field, 207$\pm$101, is one of $\sim1/4$ of  the \bgq\ results that are significantly greater than \bgg, 
while the \bgq\ result from the 90Prime $r$ band image is negative but consistent with \bgg, and consistent with
the value found by \cite{Finn2001} from {\it HST}/WFPC2 F606W data within the uncertainties.

\subsection{PG1116+215, $z=0.177$}
PSF problems affected the deep H band imaging of the host galaxy of this radio quiet object, with some conflicting results.
\cite{Guyon2006} find that the host is disk-like and asymmetric with some evidence for a disturbance, while
\cite{Veilleux2009} find that it shows no sprial arms or bar or morphological disturbance  and classify it as an old merger, in agreement
with the optical imaging \citep{Bahcall1997}.

There are $\sim$68 galaxy clusters, groups, or candidates within this
field around PG1116+215.  The two with the smallest separations from the quasar (0.28\arcmin\ and 1.92\arcmin) are ZwCl 1116.5+2136 and
SDSSCGB 16808 \citep{McConnachie2009}, both with unknown redshifts.  
However, one cluster found in the SDSS data, MaxBCG J169.85817+21.20845, lies at 7.9\arcmin\ 
separation and has a redshift marginally consistent with PG1116+215, $z_{\rm phot}=0.170$.  The relatively
large projected and redshift separations, $\sim$1 \hinv\ Mpc, and $\sim$1630 \kms, place it toward the 
periphery of this cluster if it is
associated with it at all.

Our results for \bgq\ from the 90Prime $r$ band image, 186$\pm$134, is 
consistent with \cite{McLure2001} and \cite{Finn2001} within the large
statistical uncertainties.  The \bgq\ values found in the $u$ and $g$ filters
are also consistent with \bgg, strengthening the 
conclusion that PG1116+215 lies in the outer regions or even outside the cluster potential of
MaxBCG J169.85817+21.20845. 
However, the result for the $i$ band, 863$\pm$461, presents a contradiction to this picture as it is 
greater than \bgg, more consistent with an Abell class 1 environment \citep{McLure2001}.

\subsection{PG1307+085, $z=0.155$}
The host galaxy of this radio quiet quasar is an elliptical with no evidence of a disturbance \citep{Bahcall1997,Hamilton2002,Guyon2006,
Veilleux2009}.
The center of the $z=0.14$ cluster GMBCG J197.43444+08.33445 \citep{Koester2007} lies at a projected separation of only 43.5\arcsec\ 
from the quasar.  Given the redshift separation it is unlikely that PG1307+085 is a cluster member but many galaxies in the 90Prime field
likely are.

Our results for \bgq\ in the $r$ band are consistent with  \cite{McLure2001} and \cite{Finn2001} within the large uncertainties.
The $i$ band result for \bgq\ for this field is very large, 1461$\pm$778.
The \bgq\ value found for the $g$ band, 259$\pm$120, is also large relative to other filters.
The presence of  GMBCG J197.43444+08.33445 in the foreground may be influencing the galaxy counts for this field, though it is 
not clear why this should only be the case for the $g$ and $i$ filters and not $u$ and $r$, which both show \bgq\ consistent with \bgg.

\subsection{PG1404+226, $z=0.098$, NLS1}
\cite{Ho2005} conclude that the host galaxy of this quasar has a very low
star formation rate, $<$0.14 M$_{\odot}$ yr$^{-1}$, based on the [O~{\sc ii}] measurement of \cite{Kuraszkiewicz2000}.
It also has a low black hole mass, $\log(\rm M_{\rm BH}/\rm M_{\odot})=6.88$,
as estimated from the H$\beta$ line width 
\citep{Vestergaard2006}.
\cite{McLeod1994a} find that an exponential profile with a scale length of 
2.98 \hinv\ kpc gives a reasonable fit to their H band image of this AGN.

There are $\sim$50 clusters, groups, or candidates identified in the 50\arcmin\ around PG1404+226, including five compact groups of
\cite{McConnachie2009} with separations less than $\sim$22\arcmin\ and unknown redshift. The cluster WHL J140609.9+221605 has a separation of
8.17\arcmin\ from the quasar and a redshift of $z=0.393$.

All values of \bgq\ for this field are consistent with \bgg, except in the $g$ band, where we find \bgq =942$\pm$522.

\subsection{PG1444+407, $z=0.267$}
The morphology of the host galaxy of this radio quiet AGN is somewhat ambiguous. \cite{Bahcall1997} confirm the possible presence of a bar
suggested by \cite{Hutchings1992} as the result of an old merger stage.
Interestingly,  based on the [O~{\sc ii}] measurement of \cite{Baldwin1989}, \cite{Ho2005} infer a fairly high 
star formation rate in the host of 
19.4 M$_{\odot}$ yr$^{-1}$, on the order of that found for interacting and luminous IR galaxies.

There are 48 galaxy clusters, groups, or candidates within 50\arcmin\ of the quasar, with the closest, ZwCl 1443.8+4043, lying at
a separation of 12.6\arcmin, but unknown redshift.

For this field we find \bgq\ values in the $g$, $r$, and $i$ filters consistent with \bgg. The $r$ band result is consistent with 
small positive values with large statistical uncertainties from the other studies of \cite{McLure2001} and \cite{Finn2001}.
In the $u$ band, we find a \bgq\ $<$ \bgg\ by more than $2\sigma$, reflecting the general trend for the $u$ band to show the smallest
values of \bgq, though we note that the limiting $u$ magnitude for this field corresponds to M$^{*}$+1.4 at the redshift of
this AGN rather than to M$^{*}$+2.

\subsection{PG1545+210 (3C 323.1), $z=0.264$}
This FRII radio loud AGN has an elliptical host galaxy \citep{McLeod1994,Bahcall1997} with a faint elliptical companion galaxy at a projected 
separation of 7.7 \hinv\ kpc.
\cite{Ho2005} finds a moderately large star formation rate of 9.7 M$_{\odot}$ yr$^{-1}$ in the host galaxy, 
based on the [O~{\sc ii}] measurement of \cite{Wills1993}.
From  population synthesis model fits 
to the optical spectrum of the companion,
\cite{Canalizo1997} report evidence of a possible interaction 2.3 Gyr ago.
This quasar was found to have a rich cluster environment (Abell class 1) by \cite{Oemler1972}, and
\cite{Kirhakos1994} find 22 likely galaxies in their optical images described above in Sec.~\ref{sec-mrk586}.
It is located 1.96\arcmin\ away from another broad line AGN at a very similar redshift (SDSS J154749.70+205056.5, $z=0.26547$).

These AGNs are likely  members of the X-ray indentified cluster ZwCl 1545.1+2104 ($z=0.266$, \cite{Piffaretti2011}),
also found in the SDSS DR7 data ($z=0.26735$, \cite{Hao2010}).  There are $\sim$70 other galaxy clusters, groups, or candidates within
50\arcmin\ of PG1545+210. We note that this field also contains the highest number of $z<0.5$ AGN 
of any of the other 90Prime fields, listed in Table~\ref{tab-allq}.

Here we a \bgq\ value in $g$ consistent with \bgg, and the $r$ band value identically zero, as for MRK 586.
This is in agreement with the {\it HST}/WFPC2 F606W results 
of \cite{McLure2001}.
However, given the smaller uncertainty quoted  by \cite{Finn2001}, our value $r$ band value 
differs from their positive results by $\sim3\sigma$.
As for PG1307+085, we find a very large value of \bgq\ for this field in the $i$ band and a much smaller but significant
signal in $u$, consistent with its known
cluster environment.  The limiting $u$ magnitude for this field corresponds to M$^{*}$+1.7 at the redshift of this AGN,
rather than reaching to M$^{*}$+2.

\subsection{PG1612+261, $z=0.131$}
\cite{McLeod1994a} find that an exponential profile with a scale length of
1.16 \hinv\ kpc gives a reasonable fit to their H band image of this AGN and its immediate environment meets
the criteria of \cite{McLeod1994} for bright interacting companion. (See Sec.~\ref{sec-pg0844}.)

It may be associated with the cluster NSC J161419+260832 ($z=0.1396$) but given the velocity separation and the
$\sim$0.5 \hinv\ Mpc separation on the sky from the reported cluster center, it is likely on the periphery. 
There are $\sim$47 other unique galaxy clusters, groups or candidates within 50\arcmin\ of the quasar found in NED.

The $g$ band \bgq\ value for this field, 354$\pm$237, is marginally greater than \bgg,
but it is consistent with \bgg\ in $u$, $r$ and $i$.
The most closely separated $z<1$ QSO in this field, SDSS J161335.33+263127.7, lies at $z=0.708$ and $\sim$28.5\arcmin.

\subsection{[HB89] 2141+175, $z=0.211$}
This radio loud AGN has an elliptical host, with some suggestion of an interacting companion
\citep{Hutchings1992,Dunlop2003}.
There is only one known galaxy cluster in this field, within 46\arcmin\ of this quasar.

We have no $i$ band image from 90Prime to compare with the {\it HST}/WFPC2 F675W result of \cite{McLure2001}, 
but our large $r$ band result, 991$\pm$355 is larger than theirs by $>$2.5$\sigma$.

\bibliography{jscott}
\bibliographystyle{apj}

\end{document}